\documentclass{article}
\usepackage{amsmath,amssymb,amsfonts,amsthm,xcolor}
\usepackage{graphicx,float}
\usepackage{paralist}
\usepackage{hyperref}
\usepackage{caption,subcaption}
\usepackage{siunitx,tabularx}

\usepackage{dcolumn,booktabs,paralist,float}

\usepackage[round]{natbib}
\title{Scalable logistic regression with crossed random effects}
\author{Swarnadip Ghosh\\Stanford University
  \and Trevor Hastie\\Stanford University
\and Art B. Owen \\Stanford University}
\date{December 2021}

\allowdisplaybreaks

\newtheorem{lemma}{Lemma}
\newtheorem{theorem}{Theorem}
\newtheorem{proposition}{Proposition}

\newcommand{\bone}{\mathbf{1}}

\newcommand{\mrd}{\mathrm{d}}
\newcommand{\tr}{\mathrm{tr}}
\newcommand{\ql}{\mathrm{ql}}
\newcommand{\pl}{\mathrm{pl}}

\newcommand{\logistic}{\mathrm{Logistic}}
\newcommand{\LR}{\mathrm{LR}}

\newcommand{\glmm}{\mathrm{GLMM}}

\newcommand{\logit}{\mathrm{logit}}

\newcommand{\tran}{\mathsf{T}}
\newcommand{\wh}{\widehat}

\newcommand{\diag}{\mathrm{diag}}

\renewcommand{\ge}{\geqslant}
\renewcommand{\le}{\leqslant}
\newcommand{\rd}{\,\mathrm{d}}

\newcommand{\dnorm}{\mathcal{N}}

\newcommand{\sumdot}{\text{\tiny$\bullet$}}
\newcommand{\e}{\mathbb{E}}

\newcommand{\zij}{Z_{ij}}

\newcommand{\zsj}{Z_{sj}}

\newcommand{\nid}{N_{i\sumdot}}

\newcommand{\ndj}{N_{\sumdot j}}

\newcommand{\wij}{W_{ij}}
\newcommand{\wsj}{W_{sj}}

\newcommand{\wid}{W_{i\sumdot}}
\newcommand{\wsd}{W_{s\sumdot}}
\newcommand{\wdj}{W_{\sumdot j}}

\newcommand{\ai}{a_i}

\newcommand{\bj}{b_{j}}

\newcommand{\eij}{e_{ij}}

\newcommand{\ssa}{\sigma^2_A}
\newcommand{\ssb}{\sigma^2_B}

\newcommand{\ssai}{\sigma^{-2}_A}
\newcommand{\ssbi}{\sigma^{-2}_B}

\newcommand{\ssah}{\hat\sigma^2_A}
\newcommand{\ssbh}{\hat\sigma^2_B}

\newcommand{\ssiha}{\hat{\sigma}^{-2}_{A}}
\newcommand{\ssihb}{\hat{\sigma}^{-2}_{B}}

\newcommand{\yij}{Y_{ij}}

\newcommand{\pij}{p_{ij}}

\newcommand{\xij}{\bsx_{ij}}

\newcommand{\muij}{\mu_{ij}}

\newcommand{\var}{\mathrm{var}}    
\newcommand{\cov}{\mathrm{cov}}   

\newcommand{\simiid}{\stackrel{\mathrm{iid}}{\sim}}
\newcommand{\simind}{\stackrel{\mathrm{ind}}{\sim}}

\newcommand{\real}{\mathbb{R}}

\newcommand{\phe}{\phantom{=}\,\,} 

\newcommand{\bsa}{\boldsymbol{a}}
\newcommand{\bsb}{\boldsymbol{b}}

\newcommand{\bsw}{\boldsymbol{w}}
\newcommand{\bszero}{\boldsymbol{0}}

\newcommand{\bsx}{\boldsymbol{x}}

\newcommand{\dbern}{\mathrm{Bern}}

\newcommand{\pup}{\Upsilon}

\newcommand{\cS}{\mathcal{S}}
\newcommand{\genr}{r}

\newcommand{\cs}{\mathcal{S}}

\newcommand{\cw}{\mathcal{W}}
\newcommand{\cx}{\mathcal{X}}
\newcommand{\cy}{\mathcal{Y}}
\newcommand{\cz}{\mathcal{Z}}

\newcommand{\cza}{\mathcal{Z}_A}
\newcommand{\czb}{\mathcal{Z}_B}

\newcommand{\ci}{\mathcal{I}}

\newcommand{\giv}{\!\mid\!}

\newcommand{\omlo}{\underline\omega}

\newcommand{\err}{\mathrm{Err}}
\newcommand{\dev}{\mathrm{dev}}

\begin{document}
\maketitle

\begin{abstract}
The cost of both generalized least squares (GLS) and Gibbs sampling in a crossed
random effects model can easily grow faster than $N^{3/2}$ for $N$ observations. \cite{ghos:hast:owen:2021} develop a backfitting algorithm that reduces
the cost to $O(N)$.
Here we extend that method to
a generalized linear mixed model for logistic regression.
We use backfitting within an iteratively
reweighted penalized least squares algorithm.
The specific approach is a version of penalized quasi-likelihood
due to \cite{scha:1991}. A straightforward version of Schall's algorithm
would also cost more than $N^{3/2}$ because it requires the
trace of the inverse of a large matrix.
We approximate that quantity at cost $O(N)$
and prove that this substitution makes an asymptotically negligible difference.
Our backfitting
algorithm also collapses the fixed effect with one random effect at a time
in a way that is analogous to the collapsed
Gibbs sampler of \cite{papa:robe:zane:2020}.
We use a symmetric operator that facilitates efficient covariance computation.
We illustrate our method on a real dataset from Stitch Fix.
By properly accounting for crossed random
effects we show that a naive logistic
regression could underestimate sampling
variances by several hundred fold.
\end{abstract}


\section{Introduction}

Crossed random effects structures are ubiquitous in science,
engineering and commerce.  A biologist might study a response
across many genotypes and environments \citep{bolk:etal:2009}. A social scientist
might study test scores by question and by student
\citep{baayen2008mixed}.
A web store might study responses of customers to various products.
In all of these cases we have a pair of categorical
variables with a potentially large number of levels.  When
those levels are a subset of a much larger set of potential
levels then it is natural to model them as random effects.
The problems we are interested in are large and sparse.
There are $R$ (for rows)
levels of one factor crossed with $C$ (for columns)
of another factor.  The total number of observations is $N$
and sparsity means that  $N\ll RC$.
The pattern of observations is also unstructured.

While large online commerce data sets have the crossed random effects
structure, it is exceedingly difficult to analyze such data that way.
Even for regression problems, the cost of
computing the Gauss-Markov generalized least squares estimator
grows as fast as $N^{3/2}$ or worse.
The same holds for the cost of evaluating a Gaussian likelihood even once.
The fundamental cause is that a linear equation of size
$R+C$ must be solved.  The usual algorithms solve it at cost
proportional to $(R+C)^3$
and then because $RC\ge N$ we have $\max(R,C)\ge\sqrt{N}$
and now $(R+C)^3 = \Omega(N^{3/2})$.
See \cite{crelin} for details.

A cost of $N^{3/2}$ is infeasible in  big data settings where
$N$ is very large and growing.
It is only recently that algorithms with an $O(N)$ cost
have been produced.
\cite{ghos:hast:owen:2021}
develop a backfitting algorithm
based on the work of~\cite{buja:hast:tibs:1989}.
That algorithm computes the generalized least squares
estimate and they give conditions where only $O(1)$ iterations are needed, with
each iteration taking $O(N)$ work.
A plain Gibbs sampler has the same $\Omega(N^{3/2})$ cost as
naive linear modeling \citep{crevarcomp}.
A collapsed Gibbs sampler \cite{papa:robe:zane:2020}
can attain a cost of $O(N)$ under a strong balance condition
that \cite{ghos:zhon:2021:tr} show can be weakened.

The response in an electronic commerce setting is usually
categorical and very often binary, instead of the
real valued responses that are common in other settings,
such as agriculture.
In a fixed effects setting, extending from least squares
to generalized least squares is often done by a straightforward
iteratively reweighted least squares approach. In a random
effects setting, this path is not so simple because there is
a very high dimensional integration problem that complicates
it.  Our main contribution is to find a way to adapt a
penalized quasi-likelihood algorithm of \cite{scha:1991}
to mixed effects logistic regression models where
there are crossed random effects.
Schall's algorithm as written would cost $\Omega(N^{3/2})$
making it infeasible. We have adapted it by
ignoring some off diagonal blocks in an $(R+C)\times(R+C)$
matrix and then showing that the effect of this simplification
is asymptotically negligible.
A second part of this adaptation comes from the fact
that many of the feature variables in the commerce
problem are functions of the product alone or of the
customer alone.  That brings a linear dependence that
can slow convergence of iterative methods that
alternate between updating random effects and fixed
effects.  We have devised a `clubbing' algorithm that
makes simultaneous updates to avoid this problem
as described in more detail below.

We model a binary variable $\yij\in\{0,1\}$
in terms of covariates $\xij\in\real^p$
and row and column random effects $\ai$ and $\bj$. We assume that $p$ does not grow with $N$. We leave $p$ out of our cost estimates, giving the complexity in $N$.
Conditionally on $\bsa=(a_1,\dots,a_R)^\tran$
and $\bsb=(b_1,\dots,b_C)^\tran$ the $\yij$
are independent with
\begin{align}\label{eq:basicmodel}
\Pr(\yij=1\giv \bsa,\bsb ) =
\Pr(\yij=1\giv \ai,\bj ) =
\frac{\exp(\xij^\tran\beta+\ai+\bj)}
{1+\exp(\xij^\tran\beta+\ai+\bj)}.
\end{align}
The random effects are
$\bsa\sim\dnorm(0,\ssa I_R)$
independently of $\bsb\sim\dnorm(0,\ssb I_C)$.
In this model the $\xij$ are nonrandom, either
because they were designed, or more usually because
our analysis is conditional on their observed values.

We can also write~\eqref{eq:basicmodel} as
\begin{align}\label{eq:latentmodel}
\yij =
\begin{cases}1, & \xij^\tran\beta+\ai+\bj+\eij>0\\
0,& \text{else}
\end{cases}
\end{align}
for independent random variables $\eij$ with the logistic CDF,
\begin{align}\label{eq:logisticcdf}
\Pr(\eij\le w) = \pi(w)\equiv
\frac{e^w}{1+e^w}.
\end{align}
The model~\eqref{eq:basicmodel} is a generalized linear
mixed model (GLMM) owing to the appearance of both
fixed effects $\xij^\tran\beta$ and random effects $\ai,\bj$.

While the linear model
$\yij=\xij^\tran\beta+\ai+\bj+\eij$ can be handled
by the method of moments without assuming a distributional
form for $\ai$, $\bj$ and $\eij$, estimation for the model~\eqref{eq:latentmodel}
depends on the shape  of the distributions for $\ai$ and $\bj$
as well as $\eij$.
There is a near consensus in the literature that
GLMM models are robust to mild departures from Gaussianity
and correspondingly that such departures are hard to detect.
See \cite{mccu:neuh:2011} for references and also some
contrary points of view.
It is our view that the Gaussian model is of central importance
and the problem is already hard enough under Gaussianity to
motivate using that assumption.

The conditional likelihood of $\beta$ given $\bsa$
and $\bsb$ is
\begin{align}\label{eq:condlikelihood}
L(\beta\giv\bsa,\bsb)
&=\prod_{(i,j)}
\pi( \xij^\tran\beta+\ai+\bj)^{\yij}(1-\pi( \xij^\tran\beta+\ai+\bj))^{1-\yij}\notag\\
&=
\prod_{(i,j)}
\frac{e^{(\xij^\tran\beta+a_i+b_j)\yij}}{1+e^{\xij^\tran\beta+a_i+b_j}}
\end{align}
where the product is taken over pairs $(i,j)$ for
which $(\xij,\yij)$ is observed.
The full likelihood incorporating random effect parameters is
\begin{align}\label{eq:likelihood}
L(\beta,\ssa,\ssb)
&=\int_{\real^{R+C}}
L(\beta\giv\bsa,\bsb)
\prod_{i=1}^R\frac1{\sigma_A}\varphi\Bigl( \frac{\ai}{\sigma_A}\Bigr)
\prod_{j=1}^C\frac1{\sigma_B}\varphi\Bigl( \frac{\bj}{\sigma_B}\Bigr)
\rd\bsa \rd\bsb
\end{align}
where $\bsa\in\real^R$ and $\bsb\in\real^C$ are vectors
with components $\ai$ and $\bj$ respectively
and $\varphi(\cdot)$ is the $\dnorm(0,1)$ probability density function.

The high dimensional integral in~\eqref{eq:likelihood}
presents a major difficulty to finding estimates and confidence
intervals for $\beta$, making the generalized linear mixed model~\eqref{eq:basicmodel} much harder to work with than Gaussian linear mixed models for
regression studied in \cite{ghos:hast:owen:2021}.
The integral is daunting not only because of its dimension but also
because $L(\beta\giv\bsa,\bsb)$ is a product of $N$
probabilities and so it may easily underflow numerically.

The most popular way to handle the integral is via Laplace's method,
which uses a single point in $\real^{R+C}$ to represent
the integral, as  \cite{breslow93:_approx} do in their penalized
quasi-likelihood approach.
Higher order Laplace integration methods cost $k^{R+C}$
for an integer $k\ge1$, which is  completely infeasible for
any $k>1$ in our context.

We have chosen to work with the quasi-likelihood
formulation of~\cite{scha:1991}.
That algorithm is an iteratively reweighted penalized least squares solution
that uses some weighted least squares
fits to a working response vector in order to optimize
the Laplace approximation of the integral.
The estimand in Schall's setup is a posterior
mode of $\beta$ under a diffuse prior that we describe
below.  It is thought to be close to
the MLE except when some $\Pr(\yij=1\giv\ai,\bj)$
are very close to zero or one.
Schall's algorithm is an iterative procedure with each iteration
having two steps. Both of the steps have a cost that grows like
$(R+C)^3=\Omega(N^{3/2})$.
To get around this we use backfitting for the first step
and an approximation to the
trace of the inverse of an $(R+C)\times (R+C)$ matrix
getting an algorithm that is $O(N)$ per iteration.
We then study the impact of this approximation
and find conditions under which is asymptotically negligible.

One of the main difficulties in random effects estimation
arises from the fact that the intercept is hard to estimate.
The easiest way to see this is to consider a balanced Gaussian
model with $\yij=\mu+\ai+\bj+\eij$. We would estimate $\mu$
by the average of all $RC$  observations getting
variance $O(1/R+1/C+1/N)$ which is much larger than
the $1/N$ rate from settings with $N$ IID observations.

The intercept remains difficult in the unbalanced setting
with a binary response.
The intercept contributes $\mu\bone_N$
to the $N$-vector of logits of $\Pr(\yij=1\giv\bsa,\bsb)$.
The vector space spanned by indicators of the $\ai$
includes $\mu\bone_N$.  So does the one spanned
by the indicators of $\bj$.
This overlap among spaces slows the convergence
of iterative algorithms to solve the quasi-likelihood equations.
That motivated \cite{ghos:hast:owen:2021} to develop
a generalized least squares algorithm that alternates between updating
$\ai$ along with the intercept and updating $\bj$
also with the intercept.

A related issue arises when one of the predictor variables
in $\xij$, say $x_{ij\ell}$ is a binary variable whose value
depends on only the row index $i$. That is $x_{ij\ell} = x_{i\sumdot \ell}$
for all $j$.
For instance one of the
features in e-commerce might be a property of the customer
or of the product while a feature in agriculture
might be a property of the cultivar of wheat or
a property of the environment in which it is grown.
In such cases there is an
overlap of vector spaces like the one described above
for the intercept.
The penalized likelhood estimate for the
corresponding \emph{subset} of $\bsa$ must sum to zero. This is
true for every such partially aliased column $x_{ij\ell}$, so there
can be many different such summation constraints.
Similarly to the way the intercept is handled, the coefficient $\beta_\ell$
can then be efficiently updated \emph{together} with $\ai$, a
procedure we call \emph{clubbing}.
Clearly the same
problem can happen with variables $x_{ij\ell}$ that are categorical
with more than two levels or with variables whose level is
defined by the index $j$.  Our algorithm clubs together all $p$
variables in $\beta$
with $\bsa$ when updating $\bsa$,  and also clubs $\beta$ with $\bsb$
when updating $\bsb$.
The algorithm analyzed in \cite{ghos:hast:owen:2021} only clubs
the intercept together with the random effects.
The impetus for clubbing is that we found it made a much
greater difference in the binary regression case than
we saw in the generalized least squares setting.

An outline of this paper is as follows.
Section~\ref{sec:missing} introduces our notation
and describes the missingness mechanism.
Section~\ref{sec:howto} presents a penalized quasi-likelihood
method based on a Laplace approximation.  The optimization
is done using Schall's method after incorporating backfitting
and the matrix approximation referred to above.  We then
give conditions under which that approximation makes
an asymptotically negligible difference.
Section~\ref{sec:clubbing} describes in detail our clubbing
strategy where we alternate between updating $(\hat\beta,\hat\bsa)$
and $(\hat\beta,\hat\bsb)$.
Section~\ref{sec:empiricalresults} uses some simulated
data to verify that the cost of a standard algorithm (glmer from \cite{lme4})
has a cost growing faster than $N^{3/2}$.  Two versions of that
algorithm show the expected superlinear cost and that superlinearity
applies even to a single iteration.
Our iteratively reweighted
backfitting algorithm converges in $O(N)$ cost there.
This holds in our example even though weighting
will ordinarily violate the sufficient conditions
that \cite{ghos:hast:owen:2021} give for backfitting to
cost $O(N)$.
We also include a simulation that compares the accuracy of
glmer, and our proposal and a naive logistic regression that ignores the random effects.
Our algorithm's accuracy lies between that of two glmer algorithms,
both of which have superlinear cost. A naive logistic regression is seen
to be typically inconsistent.  For glmer and our algorithm
the error in the intercept coefficient
decreases much more slowly than that of the other coefficients,
in line with our earlier remarks.
Section~\ref{sec:stitchfix} illustrates our algorithm on a data set provided by Stitch Fix.
Section~\ref{sec:discussion} has a discussion about the Bayesian and frequentist perspectives.
In Appendix~\ref{sec:appendix}, we give details about the proofs and tabulate the results
for the Stitch Fix data.

We conclude this section by describing some alternative approaches.
Quasi-Monte Carlo (QMC) sampling \citep{nied:1992,dick:kuo:sloa:2013} is very
well suited to many high dimensional integration problems but
the integrand in~\eqref{eq:likelihood} does not appear to be one
of them.  This integrand is very `spiky' making it difficult for QMC.
See \cite{kuo:etal:2008} for some work in this area.
Composite likelihoods \citep{vari:reid:firt:2011} lower the dimension
of the integration problem by multiplying likelihoods based on
selected pairs or $k$-tuples of observations instead of all $N$ of them at once.
A comparison with composite likelihood would require consideration
of precisely how to choose the $k$-tuples as well as the resulting
efficiency.  That would require an investigation of its own and is
outside the scope of this paper.
We have opted to build upon the penalized quasi-likelihood
method of \cite{breslow93:_approx}.  It is the most commonly
used algorithm and there is well tested code in \cite{lme4}
to use in comparisons.

It would be of great interest to handle mixed and generalized linear mixed models that incorporate SVD-like latent variable interactions.  For the present we note that even without those interactions it is a significant challenge to get an efficient estimate of $\beta$ and also quantify the uncertainty in $\beta$ whether by Bayesian or frequentist methods while remaining within an $O(N)$ cost limit.

\section{Notation and sampling models}\label{sec:missing}

We speak of rows and columns for our two factors.
The rows are indexed by $i=1,\dots,R$ and the
columns by $j=1,\dots,C$.  Most of the $RC$ possible $(i,j)$ combinations
are not observed in our motivating problems.
We let $\zij\in\{0,1\}$ take the value $1$ if and only if
$(\xij,\yij)$ has been observed. In our motivating problems
we either never have the same $(i,j)$ combination observed
twice, or we only keep the most recent such observation,
or it happens so rarely that we can neglect it.

Sums over $i$ are from $1$ to $R$
and sums over $j$ are from $1$ to $C$.
The number of times row $i$ has been observed is
$\nid = \sum_j\zij$ and similarly, column $j$ has been
observed $\ndj=\sum_i\zij$ times.
We suppose that the data are ordered so that all $\nid\ge1$
and all $\ndj\ge1$.
The total sample size is $N=\sum_{i=1}^{R}\nid=\sum_{j=1}^{C}\ndj$. The sparsity condition implies that $N \ll RC$.

Sometimes we have to make a vector of length $N$ with an element
for each observation or a matrix with $N$ rows, one per observation.
We assume that some consistent ordering of the
observations is used in all of these cases
and we use a calligraphic font for most such quantities.
The vector $\cy\in\{0,1\}^N$ contains all
the observed responses $\yij$.
The matrix $\cza\in\{0,1\}^{N\times R}$  has $R$ columns
of which the $i$'th column contains ones
for observations in row $i$ and zeros everywhere else.
The matrix $\czb\in\{0,1\}^{N\times C}$ is the corresponding
incidence matrix for the columns of our observations.
The product $\cza^\tran \czb$ is our observation matrix $Z\in\{0,1\}^{R\times C}$.

While the analyst considers $\zij$ to be fixed, we
will study random $\zij$ in order to model
the difficulties that the analyst will face
and to prove that certain difficulties have vanishing
probability in large samples.
The model we use comes from \cite{ghos:hast:owen:2021}.
For a problem size parameter $S$, the number of rows
and columns are $R=\lfloor S^\rho\rfloor$ and $C=\lfloor S^\kappa\rfloor$
respectively.  Those values become very large in cases of interest
and, to avoid uninteresting complications, we use $R=S^\rho$ and $C=S^\kappa$
in our formulas without taking integer parts.
Then for some $\Upsilon>1$, we suppose that
\begin{align}\label{eq:zmodel}
\zij \simind \dbern(\pij)\quad
\text{where}\quad
 \frac{S}{RC}\le \pij\le\Upsilon\frac{S}{RC}.
\end{align}
The actual problem size $N$ satisfies
$S\le \e(N)\le\Upsilon S$.
The model~\eqref{eq:zmodel} allows unequal
observation probabilities but remains
more restrictive than we would like.
Numerical results in \cite{ghos:hast:owen:2021}
show that their generalized least squares algorithm
converges in $O(N)$ cost much more generally
than their assumptions imply.

The relevant values of $\rho$ and $\kappa$
are positive so that $R$ and $C$ grow with $N$
but below $1$ because neither $R$ nor $C$ can
be larger than $N$.
We also have $\rho+\kappa>1$ to model
sparsity, that is $N\ll RC$.

Under the model~\eqref{eq:zmodel},
\cite{ghos:hast:owen:2021} show that
if $2\rho+\kappa<2$ then all rows get an adequate
sample size and no single row dominates the data:
\begin{align}\label{eq:gotnid}
\lim_{S\to\infty} \Pr\bigl(
(1-\epsilon)S^{1-\rho} \le\min_i\nid \le\max_i\nid\le(\Upsilon+\epsilon)S^{1-\rho}
\bigr) &= 1
\end{align}
for any $\epsilon>0$.
By symmetry, if $\rho+2\kappa<2$, then
\begin{align}\label{eq:gotndj}
\lim_{S\to\infty} \Pr\bigl(
(1-\epsilon)S^{1-\kappa} \le\min_j\ndj \le\max_j\ndj\le(\Upsilon+\epsilon)S^{1-\kappa}
\bigr) &= 1
\end{align}
holds for any $\epsilon>0$.

Many settings can be expected to have informative
missingness, where $\zij$ is related to $\yij$.
We do not account for this because to do so would
require problem specific assumptions from outside the
data at hand.  Also, the problems we face are already
unsolved in the event that $\zij$ are
not informative about~$\yij$.

\section{Likelihood approach}\label{sec:howto}
We will use an iterative approach of \cite{scha:1991}.
\cite{breslow93:_approx} showed the resulting estimators obtained by
this way jointly maximize (under very mild assumptions)
the penalized quasi-likelihood of \cite{green:1987}
and also maximize the Bayesian posterior probability density function.
It alternates between estimating $\bsa$, $\bsb$
and $\beta$ given $\ssa$, $\ssb$ and an overdispersion
quantity $\phi$ and estimating the variances and overdispersion
given $\bsa$, $\bsb$ and $\beta$.
For the optimization
over $\bsa$ and $\bsb$ and $\beta$
we use weighted least squares
for a working response, following the
approach in \cite{scha:1991}.
His approach requires the trace of the inverse
of an $(R+C)\times(R+C)$ matrix and this
is too expensive so we replace it by an
approximation. We then give conditions
where the approximation makes an asymptotically
negligible difference.



We observe $\yij$ together with predictor variables $\xij$.
There are also unobserved random effects $(\ai, \bj)$  associated with
this observation.
We assumed that given $\bsa$ and $\bsb$, the $\yij$ are conditionally
independent.
Adapting some notation from \cite{breslow93:_approx}
we let $\mu^{\bsa,\bsb}\in\real^N$
have components
$\muij^{\bsa,\bsb} \equiv \e(\yij\giv\ai,\bj)= \pi(\xij^\tran\beta+\ai+\bj)$.
The model of \cite{breslow93:_approx}
includes
$$\var(\yij \giv \ai, \bj) = \phi v(\mu_{ij})$$
for a function $v(\cdot)$ and an overdispersion parameter $\phi>0$.
In our case $v(\mu)=\mu(1-\mu)$.
While Bernoulli random variables cannot be overdispersed or
underdispersed we include $\phi$ because
values $\phi\ne1$ may signify a lack of fit.
The integrated quasi-likelihood $\ql(\beta,\ssa,\ssb,\phi)$ from
 \cite{breslow93:_approx} is
\begin{align}
e^{\ql(\beta,\ssa,\ssb,\phi)}
&\propto\sigma_{A}^{-R} \sigma_{B}^{-C}\int_{\real^{C}} \int_{\real^{R}}
e^{-\gamma(\bsa,\bsb)}\rd\bsa \rd\bsb,\quad\text{where} \nonumber\\
\gamma(\bsa,\bsb)&=
\frac1{2\phi}\sum_{ij} \zij \dev(y_{ij} ,\mu_{ij}^{\bsa,\bsb}) + \frac{\Vert\bsa\Vert^2}{2\ssa} +\frac{\Vert\bsb\Vert^2}{2\ssb} \label{eq:greenpql}
\end{align}
for a Bernoulli deviance
$$
\dev(y, \mu)= -2y\log(\mu)-2(1-y)\log(1-\mu)
$$
taking $0\log0=0$.
The variables $\xij$, $\yij$, $\zij$, $\sigma_A$, $\sigma_B$, $\phi$ and $\beta$ are absorbed into the definition of $\gamma$
so that we can focus on the variables $\bsa$ and $\bsb$ over which we integrate.

We are going to use Laplace's method of approximate integration.
For that we need the Hessian $\gamma''$ of $\gamma$ at the point $(\tilde\bsa,\tilde\bsb)$ where
its gradient $\gamma'$ vanishes.
Applying Laplace's method of integral approximation we get
$$\ql(\beta,\ssa,\ssb,\phi) \approx-R\log\sigma_A-C \log\sigma_B
-\frac{1}{2} \log \det(\gamma''(\tilde{\bsa},\tilde{\bsb}))-\gamma(\tilde{\bsa},\tilde{\bsb}).$$
The gradient in equation (8) of \cite{breslow93:_approx}
involves the quantity $v(\mu)\pi'(\mu)$ which equals $1$
in our context.  After translating their random effects model into
our context we find that
\begin{align*}
\gamma'(\bsa,\bsb) &=
\begin{pmatrix}
-\cza^\tran(\cy-\mu^{\bsa,\bsb})/\phi+\bsa/\ssa\\[1ex]
-\czb^\tran(\cy-\mu^{\bsa,\bsb})/\phi +\bsb/\ssb
\end{pmatrix}\quad\text{and}\\
\gamma''(\bsa,\bsb) &=
\begin{pmatrix}
\cza^\tran\cw\cza+\ssai I_R & \cza^\tran\cw\czb\\[1ex]
\czb^\tran\cw\cza& \czb^\tran\cw\czb+\ssbi I_C
\end{pmatrix}
\end{align*}
where $\cw=\diag( \mu^{\bsa,\bsb}(1-\mu^{\bsa,\bsb}))/\phi\in[0,1/(4\phi)]^{N\times N}$.

 Using the expressions above,
\begin{align}\label{eq:qleq}
\begin{split}
\ql(\beta,\ssa,\ssb,\phi) &\approx-\frac{1}{2}\log \det(D)
-\frac{1}{2} \log \det(\cz^\tran\cw(\mu^{\tilde{\bsa},\tilde{\bsb}})\cz + D^{-1})-\gamma(\tilde{\bsa},\tilde{\bsb})\\
&\approx-\frac{1}{2}\log \det(I + \cz^\tran\cw(\mu^{\tilde{\bsa},\tilde{\bsb}})\cz D)-\gamma(\tilde{\bsa},\tilde{\bsb})\\
&\approx-\frac{1}{2} \log \det(I + \cz^{\tran}\cw(\mu^{\tilde{\bsa},\tilde{\bsb}})\cz D)\\
& \quad\quad\quad\quad-\frac1{2\phi}\sum_{ij} \zij \dev(\yij,\mu_{ij}^{\tilde{\bsa},\tilde{\bsb}})-\frac{\Vert\tilde\bsa\Vert^2}{2\ssa}  - \frac{\Vert\tilde\bsb\Vert^{2}}{2\ssb}
\end{split}
\end{align}
for $\cz = [\cza : \czb]$ and $D = \diag(\ssa I_{R},\ssb I_{C}).$

 If $\ssa$, $\ssb$ and $\phi$ are known, we can find $\hat{\beta},\hat{\bsa}(\beta,\ssa,\ssb,\phi),\hat{\bsb}(\beta,\ssa,\ssb,\phi)$ to maximize negative $\gamma$ in~\eqref{eq:greenpql},
the penalized quasi-likelihood of \cite{green:1987}.
Following \cite{breslow93:_approx}
we assume that the GLM iterative weights vary at most slowly
as a function of the mean, and so we ignore the first term
in the expression~\eqref{eq:qleq} and choose parameters to maximize the rest.
We obtain the score equation upon differentiating with respect to
$(\beta,\bsa,\bsb)$ which gives us the following three equations:
\begin{align}\label{eq:newtonraphson0}
\begin{split}
\cx^{\tran}(\cy-\mu^{\bsa,\bsb}) &= 0,\\
\cza^{\tran}(\cy-\mu^{\bsa,\bsb}) &= \frac{1}{\ssa}\bsa,\quad\text{and}\\
\czb^{\tran}(\cy-\mu^{\bsa,\bsb}) &= \frac{1}{\ssb}\bsb.
\end{split}
\end{align}

We solve for $\hat{\beta}$ using an iterative reweighted least squares approach. Letting ${\theta} = (\beta,\ssa,\ssb,\phi)$,
the regression coefficient $\hat{\beta}^{(k+1)}$ is obtained by solving a weighted least squares problem with weights depending on $\hat{\theta}^{(k)}$. In the following section, we give a detailed description of the procedure.

Using the Hessian, we obtain  (after a bit of rearrangement) the following three equations involving a working response $z$:
\begin{align}\label{eq:newtonraphson1backfit}
\begin{split}
\cx\hat{\beta} &= \cx(\cx^{\tran}\cw\cx)^{-1}\cx^{\tran}\cw(z-\cza \hat{\bsa} - \czb \hat{\bsb}),\\
\cza\hat{\bsa} &= \cza
(\cza^{\tran}\cw\cza+\ssai I_R)^{-1}\cza^{\tran}\cw(z-\cx \hat{\beta} - \czb \hat{\bsb}),\\
\czb\hat{\bsb} &=
\czb(\czb^{\tran}\cw\czb+\ssbi I_C)^{-1}\czb^{\tran}\cw(z-\cx \hat{\beta} - \cza \hat{\bsa}).\\
\end{split}
\end{align}
Here the working response $z$ is defined
\begin{equation}
  \label{eq:workingresponse1}
 z_{ij} = \xij^\tran\hat{\beta} +\hat{a}_{i}+\hat{b}_{j} + \frac{\yij - \hat{\mu}_{ij}}{\hat{\mu}_{ij}(1-\hat{\mu}_{ij})}
\end{equation}
with $\hat\mu_{ij} = \pi( \xij^\tran\hat\beta+\hat a_i+\hat b_j)$
for the logistic CDF $\pi(\cdot)$ from equation~\eqref{eq:logisticcdf}.
A straightforward solution of equation~\eqref{eq:workingresponse1}
costs $O((R+C+p)^3)$ which is infeasible and so we
develop a backfitting iteration for it.

\subsection{Schall's approach}\label{subsec:schallapproach}

\cite{scha:1991} considers an iterative procedure to estimate the fixed effect, random effects and the variance component. The iterative procedure is based on the quasi-likelihood from the prior subsection.
At iteration $k$ we have
estimates $\hat\beta^{(k)}$, $\bsa^{(k)}$, $\bsb^{(k)}$,
$\hat\sigma^{2(k)}_A$,
$\hat\sigma^{2(k)}_B$ and $\hat\phi^{(k)}$.
For $k=0$, we
initialize them all to zeros except the
overdispersion and variance parameters start at one.
To obtain $\hat{\beta}^{(k+1)}$, $\hat{\bsa}^{(k+1)}$ and $\hat{\bsb}^{(k+1)}$
we solve the penalized weighted least squares problem
\begin{align}\label{eq:minboth}
\min_{\beta,\bsa,\bsb} \sum_{ij} \zij\,\wh{W}_{ij}^{(k)}
\bigl( z_{ij}^{(k)}-\xij^\tran\beta-\ai-\bj \bigr)^2
+\frac{\Vert\bsa\Vert^2}{\hat{\sigma}^{2(k)}_{A}} +\frac {\Vert\bsb\Vert^2}{\hat{\sigma}_{B}^{2(k)}},
\end{align}
with weights
$\wh{W}_{ij}^{(k)} =
{\hat{\mu}_{ij}^{(k)}(1-\hat{\mu}_{ij}^{(k)})}/{\hat{\phi}^{(k)}}$. This
optimization problem
leads to exactly the same estimating equations
\eqref{eq:newtonraphson1backfit}, since the quadratic approximation is
based on the same gradients and Hessian.

It remains to update the overdispersion and variance
parameters.
Consider the following $(R+C)\times(R+C)$ matrix
\begin{align}\label{eq:defczgtczg}
\begin{split}
  T^{(k)}&=
\begin{pmatrix}
T^{(k)}_{11} & T^{(k)}_{12}\\[.5ex]
T^{(k)}_{21} & T^{(k)}_{22}
\end{pmatrix}\\
&=
\begin{pmatrix}
    \cza^\tran \widehat{\cw}^{(k)} \cza+{\hat{\sigma}^{-2(k)}_{A}}{I}_{R}&\cza^\tran \widehat{\cw}^{(k)} \czb\\[0.25ex]
      \czb^\tran \widehat{\cw}^{(k)} \cza& \czb^\tran \widehat{\cw}^{(k)} \czb+{\hat{\sigma}^{-2(k)}_{B}}{I}_{C}
\end{pmatrix}
\end{split}
\end{align}
with $\wh{\cw}^{(k)} = \diag(\wh{W}_{ij}^{(k)})\in\real^{N\times N}.$
Let
$$T^{*(k)} = \bigl(T^{(k)}\bigr)^{-1} =
\begin{pmatrix}
    T^{*(k)}_{11} & T^{*(k)}_{12}\\[.5ex]
    T^{*(k)}_{21} & T^{*(k)}_{22}
    \end{pmatrix}.$$
Schall's update evaluates
$$\nu_{A}^{(k+1)} = {\tr(T^{*(k)}_{11})}/{\hat{\sigma}^{2(k)}_{A}}\quad\text{and}\quad
\nu_{B}^{(k+1)} = {\tr(T^{*(k)}_{22})}/{\hat{\sigma}^{2(k)}_{B}}$$
and then sets
\begin{align}\label{eq:newsigmas}
\hat\sigma^{2(k+1)}_A =
\frac{\Vert\hat{\bsa}^{(k+1)}\Vert^2}
{R - \nu_{A}^{(k+1)}}
\quad\text{and}\quad
\hat\sigma^{2(k+1)}_B =
\frac{\Vert\hat{\bsb}^{(k+1)}\Vert^2}
{C - \nu_{B}^{(k+1)}}.
\end{align}
From these formulas it is clear that the $\nu$ parameters
function as degrees of freedom estimates.
Finally, we let
\begin{align}\label{eq:newphi}
\hat\phi^{(k+1)}
=
\frac{\sum_{ij}\zij\hat{\mu}_{ij}^{(k+1)}(1 - \hat{\mu}_{ij}^{(k+1)})
\bigl(z_{ij}^{(k+1)}-x_{ij}^\tran\hat{\beta}^{(k+1)} - \hat{a}_{i}^{(k+1)} - \hat{b}_{j}^{(k+1)}\bigr)^{2}}
{N-p-(R-\nu_{A}^{(k+1)})-(C-\nu_{B}^{(k+1)})}
\end{align}
where $p$ is the number of parameters in
$\beta$ including the intercept.

The quantity Schall computes is almost but not
quite the maximum likelihood estimate of $\beta$.
\cite{scha:1991} shows that it is
a quantity studied by \cite{stiratelli:1984},
namely the posterior mode of $\beta$ under a diffuse
prior on $\beta$ independent of a zero-mean Gaussian prior  for
$(\bsa,\bsb)$. \cite{stiratelli:1984} mentioned different approaches like empirical Bayes for the estimation of variance components of the prior
distribution (See \cite{leonard:1975}, \cite{laird:1978}).
The update~\eqref{eq:minboth}
for $\beta$, $\bsa$ and $\bsb$
is a Fisher scoring
iteration to maximize their posterior density.
The updates~\eqref{eq:newsigmas} and~\eqref{eq:newphi}
for $\ssa$, $\ssb$ and $\phi$
are from an EM iteration to compute these
dispersion components, after
approximating the posterior distribution
of $(\bsa,\bsb)$ by a multivariate normal distribution with
the same mode and curvature as the true posterior.
In small data sets where we are able to compute the MLE
we find the estimate from Schall's algorithm is very close to the MLE.

\subsection{Modified Schall Approach}\label{subsec:modifiedschallapproach}
We want to solve the equation~\eqref{eq:minboth}.
It is instructive to begin with the case of just one factor.
Then we obtain the following optimization problem:
\begin{align}\label{eq:mina}
\min_{\beta,\bsa}\sum_{i,j} \zij \wh{W}_{ij}^{(k)} \bigl( z_{ij}^{(k)}-\xij^\tran\beta-\ai \bigr)^2
+\frac{\Vert\bsa\Vert^2}{\hat{\sigma}^{2(k)}_{A}}.
\end{align}
The normal equations from~\eqref{eq:mina} yield
\begin{align}
\bszero & = \cx^\tran \wh{\cw}^{(k)}(z^{(k)}-\cx\hat\beta-\cza\hat\bsa),\quad\text{and}\label{eq:normbeta}\\
\bszero & = \cza^\tran \wh{\cw}^{(k)}(z^{(k)}-\cx\hat\beta-\cza\hat\bsa)  
-\hat\bsa/\hat\sigma^{2(k)}_A.
\label{eq:normbsa}
\end{align}

Solving~\eqref{eq:normbsa} for $\hat\bsa$ and multiplying the solution by $\cza$ yields
\begin{align*}
\cza\hat\bsa &= \cza\bigl(\cza^\tran \wh{\cw}^{(k)} \cza + {\hat{\sigma}^{-2(k)}_{A}}I_R\bigr)^{-1}\cza^\tran \wh{\cw}^{(k)}(z^{(k)}-\cx\hat\beta)\\
&\equiv \cs_A^{(k)}(z^{(k)} - \cx\hat\beta)
\end{align*}
for an $N\times N$ ridge regression ``smoother matrix'' $\cs_A^{(k)}$.
(We have rederived one of the estimating equations in \eqref{eq:newtonraphson1backfit}.)
This smoother matrix implements weighted shrunken within-group means.
Substituting $\cza\hat\bsa$ into equation~\eqref{eq:normbeta}    
yields
\begin{equation}
  \label{eq:onefactorsolution}
\hat\beta = \bigl(\cx^\tran \wh{\cw}^{(k)}(I_N-\cs_A^{(k)})\cx\bigr)^{-1}\cx^\tran \wh{\cw}^{(k)}(I_N-\cs_A^{(k)})z^{(k)}.
\end{equation}
None of these steps takes superlinear time since
weighted shrunken within-group means cost $O(N)$
time.
Also observe that $\wh\cw^{(k)}\cs_{A}^{(k)}$ is symmetric.
We can use this fact to efficiently compute an estimated (asymptotic)
covariance
of $\hat\beta$.
With $\wh\cw_{\cS_A}^{(k)} = \wh{\cw}^{(k)}(I_N-\cS_A^{(k)})$,
we have
\begin{equation}\label{eq:covbetahat2}
\wh\cov(\hat\beta)=(\cx^\tran \wh\cw_{\cS_A}^{(k)} \cx)^{-1}\cx^\tran
\wh\cw_{\cS_A}^{(k)} \cdot\wh\Sigma\cdot \wh\cw_{\cS_A}^{(k)}\cx(\cx^\tran \wh\cw_{\cS_A}^{(k)} \cx)^{-1},
\end{equation}
with $\wh\Sigma=\hat{\sigma}^{-2(k)}_{A} \cza\cza^{\tran} +
(\wh{\cw}^{(k)})^{-1}$ the covariance of the working response.
Since $\wh\Sigma$ has the form low-rank plus diagonal, we  can
compute this covariance with  $O(N)$ computations.

With two factors we do not enjoy the same computational
simplifications.
The counterpart to equation~\eqref{eq:onefactorsolution} is
\begin{equation}\label{eq:twofactorsolution}
\hat\beta = (\cx^\tran \wh{\cw}^{(k)}(I_N-\cs_{AB}^{(k)})\cx)^{-1}\cx^\tran \wh{\cw}^{(k)}(I_N-\cs_{AB}^{(k)})z^{(k)}
\end{equation}
where
$$\cs_{AB}^{(k)} = \cz(\cz^{\tran}\wh{\cw}^{(k)}\cz+(D^{(k)})^{-1})^{-1}\cz$$
for $\cz = [\cza : \czb]$ and $D^{(k)} = \diag(\hat{\sigma}^{2(k)}_{A} I_{R},\hat{\sigma}^{2(k)}_{B} I_{C}).$
Hence we would need to invert an $(R + C) \times (R+C)$ matrix $T^{(k)}$
in~\eqref{eq:defczgtczg} to apply $\cs_{AB}^{(k)}$ and thereby
solve~\eqref{eq:twofactorsolution}, incurring a cost far greater than $O(N)$.

However, in order to solve (\ref{eq:twofactorsolution}), all we need to do is
apply the operator $\cs_{AB}^{(k)}$ to each column of $\cx$, and this
can be done more efficiently.
Consider a generic response vector $\genr$ (such as a column of $\cx$) and the optimization problem
\begin{align}\label{eq:minabr}
\min_{\bsa,\bsb}\Vert \genr-\cza\bsa-\czb\bsb\Vert_{\wh{\cw}^{(k)}}^2
+\hat{\sigma}^{-2(k)}_{A}\|\bsa\|^2+\hat{\sigma}^{-2(k)}_{B}\|\bsb\|^2.
\end{align}

It is clear that the fitted values are given by
$\widehat\genr=\cs_{AB}^{(k)}\genr$. Solving ~\eqref{eq:minabr}
leads to the following two blocks of estimating equations:
\begin{align}\label{eq:backfittwoblocks}
\begin{split}
\cza\hat\bsa &=  \cs_A^{(k)}(\genr - \czb\hat\bsb),\\
\czb\hat\bsb &=  \cs_B^{(k)}(\genr - \cza\hat\bsa).
\end{split}
\end{align}
We can solve these equations iteratively by block coordinate descent (backfitting).
This is done in parallel with $r$ being each
column of $\cx$ (separately) obtaining
$\cs^{(k)}_{AB}\cx$ at convergence.

Similar to \eqref{eq:covbetahat2}, we obtain
the covariance estimate for the two factor case.
\begin{equation}\label{eq:covbetahat2factor}
\wh\cov(\hat\beta)=(\cx^\tran \wh\cw_{\cS_{AB}}^{(k)} \cx)^{-1}\cx^\tran \wh\cw_{\cS_{AB}}^{(k)}\cdot \widehat\Sigma\cdot  \wh\cw_{\cS_{AB}}^{(k)}\cx(\cx^\tran \wh\cw_{\cS_{AB}}^{(k)} \cx)^{-1},
\end{equation}
with $\widehat\Sigma =  \hat{\sigma}^{-2(k)}_{A} \cza\cza^{\tran}
+\hat{\sigma}^{-2(k)}_{B} \czb\czb^{\tran}+ (\wh{\cw}^{(k)})^{-1}$ the
covariance of the working response.
Again, because of the low-rank-plus-diagonal nature of $\wh\Sigma$, we can
compute the covariance with  $O(N)$ computations.

In practice, we will need to repeatedly
minimize \eqref{eq:minboth} for each step $k$ as the weights
${\wh{\cw}^{(k)}}$ and $\hat{\sigma}^{-2(k)}_{A}$ and
$\hat{\sigma}^{-2(k)}_{B}$ vary, and only compute the covariance
estimate after convergence. We develop a more efficient algorithm for this purpose which we describe in Section~\ref{sec:clubbing}.

Now we consider the second step of Schall's method.
Recall that $T^{(k)}$ is a $(R+C) \times (R+C)$ matrix without any special structure.

Although the diagonal blocks in $T^{(k)}$ are diagonal matrices, the off diagonal block
have no special structure.
Inverting $T^{(k)}$ would have a cost of $N^{3/2}$ or worse.
Our approximation to Schall's algorithm uses
$$
\nu_A^{(k+1)} = \tr( (T^{(k)}_{11})^{-1})/\hat\sigma^{2(k)}_A
\quad\text{and}\quad
\nu_B^{(k+1)} = \tr( (T^{(k)}_{22})^{-1})/\hat\sigma^{2(k)}_B
$$
simply ignoring the off diagonal blocks of $T^{(k)}$.
These can be computed in $O(N)$ time.
Getting the trace of $(T^{(k)}_{11})^{-1}$ and $(T^{(k)}_{22})^{-1}$
costs $O(R+C)$ because those matrices are diagonal.
The $i$'th diagonal element of $T^{(k)}_{11}$ is
$\sum_j\zij\wh{W}_{ij}^{(k)}+\hat{\sigma}^{-2(k)}_{A}$ and so all elements
of $(T^{(k)}_{11})^{-1}$ and $(T^{(k)}_{22})^{-1}$
can be computed in $O(N)$ work.
We show in the following subsection that this approximation
ignoring the off diagonal blocks makes an asymptotically
negligible difference.

Schall's approach requires the trace for each of two blocks of the inverse of
the partitioned matrix
\begin{align}\label{eq:schallst}
T =
\begin{pmatrix}
\cza^\tran \cw \cza  +\sigma^{-2}_AI_R
 & \cza^\tran \cw \czb\\
\czb^\tran \cw \cza & \czb^\tran \cw \czb + \sigma^{-2}_BI_C
\end{pmatrix}
\end{align}
with a diagonal weight matrix $\cw\in(0,1/(4\phi)]^{N\times N}$.
Computing those traces directly costs $O((R+C)^3)$
because of the inversion step
and this is infeasible
in our applications. Instead ignoring the
off-diagonal blocks of $T$ as mentioned
above leads to our use of
\begin{align}\label{eq:cheapotrace}
\tr( (\cza^\tran \cw \cza  +\sigma^{-2}_AI_R)^{-1})
\quad\text{and}\quad
\tr( (\czb^\tran \cw \czb  +\sigma^{-2}_BI_C)^{-1})
\end{align}
for the two traces of blocks of $T^{-1}$.

We justify this approximation in two steps.
First we give a representation for the
error incurred in ignoring off diagonal
blocks when taking the trace of the inverse.
Then we show that error formula is asymptotically
negligible under a sampling model for $\zij$.
We work with the true weights $W_{ij}$ and comment
later on the implications for estimated
weights $\wh W_{ij}$.

We define the weight sums
$$\wid = \sum_j\zij\wij\quad\text{and}\quad
\wdj=\sum_i\zij\wij.$$
Then writing
$$
T =
\begin{pmatrix}
\diag(\wid + \ssai) & W\\
W^\tran & \diag(\wdj+\ssbi)
\end{pmatrix}
$$
we see that the traces in~\eqref{eq:cheapotrace}
involve diagonal matrices and so they
can be computed in $O(R+C)$ work
after $O(N)$ work to compute their elements.

For $0\le\eta\le1$ let
\begin{align}\label{eq:defteps}
T(\eta) &=
\begin{pmatrix} A & \eta B\\
\eta B^\tran & C
\end{pmatrix}
\end{align}
where matrices $A$, $B$ and $C$ are defined so that
$T(1)$ is Schall's matrix $T$ from~\eqref{eq:schallst}.
There should be no confusion between the matrix $C$
here and our number of columns.
Blocks $A$ and $C$ are diagonal in our problem and
$T(\eta)$ is diagonally dominant for all $0\le \eta\le1$.
Therefore $T=T(1)$ is invertible.
In the proof of Lemma~\ref{lem:pert} we will see
that $\tr(T(\eta)^{-1})-\tr(T(0)^{-1})=O(\eta^2)$ so that
small perturbations $\eta B$ do not bring a large
approximation error. However we need to study $\eta=1$
so a small $\eta$ analysis is not quite enough, so
the following Lemma~\ref{lem:spectralbound} shows that
$\eta=1$ is within the radius of convergence.

\begin{lemma}\label{lem:pert}
Let $T(\eta)\in\real^{(R+C)\times(R+C)}$ have the form~\eqref{eq:defteps}
for positive definite diagonal matrices $A$ and $C$.
Define $B_*=A^{-1/2}BC^{-1/2}$ and let $\rho$ be
the spectral radius of
$$\begin{pmatrix} 0 & B_*\\
B_*^\tran & 0\end{pmatrix}.$$
Then for $0\le\eta<1/\rho$,
$$
\tr(T(\eta)^{-1})=\tr(T(0)^{-1})
+\tr\biggl(A^{-1}\sum_{k\ge1} (\eta^2B_*B_*^\tran)^k\biggr)
+\tr\biggl(C^{-1}\sum_{k\ge1} (\eta^2B_*^\tran B_*)^k\biggr).
$$
\end{lemma}
\begin{proof}
See Appendix~\ref{sec:proof:lem:pert}.
\end{proof}

The above Lemma gives the error in
the trace.  From the proof we see that
the two specific error terms that we need
to bound are
\begin{align}\label{eq:errorstobound}
\tr\Bigl(A^{-1}\sum_{k\ge1}(B_*B_*^\tran)^k\Bigr)
\quad\text{and}\quad
\tr\Bigl(C^{-1}\sum_{k\ge1}(B_*^\tran B_*)^k\Bigr)
\end{align}
after setting $\eta=1$.
To use Lemma~\ref{lem:pert} with $\eta=1$ we need to verify
that the given spectral radius $\rho$
is below $1$ for our setting.
We use Lemma~\ref{lem:spectralbound}
to control the spectral radius.

\begin{lemma}\label{lem:spectralbound}
In the notation of Lemma~\ref{lem:pert},
let $B$ have {nonnegative} entries.
Then the spectral radius $\rho$ satisfies
$$
\rho \le
\biggl({\max_i\sum_j \frac{B_{ij}}{C_{jj}}}\biggr)^{1/2}
\biggl({\max_j\sum_i \frac{B_{ij}}{A_{ii}}}\biggr)^{1/2}.
$$
\end{lemma}
\begin{proof}
See Appendix~\ref{sec:proof:lem:spectralbound}.
\end{proof}

In our present context
$A=\diag(\wid+\ssai)$,
$C=\diag(\wdj+\ssbi)$
and $B=W$.
Then
$$
\max_i\sum_j\frac{B_{ij}}{C_{jj}}
=\max_i\sum_j\frac{W_{ij}}{\wid+\ssai}
=\max_i\frac{\wid}{\wid+\ssai}<1
$$
and we get
$$
\rho <
\biggl(\max_i\frac{\wid}{\wid+\ssai}
\times
\max_j\frac{\wdj}{\wdj+\ssbi}\biggr)^{1/2}<1
$$
as required.

Now $\wij = \pij(1-\pij)/\phi\in(0,1/(4\phi)]$ for all $(i,j)$ in the
data. We will need a strictly positive lower bound
$\wij\ge\omlo>0$.
In many applications, we can reasonably assume that $|\xij^\tran\beta|$
is bounded away from infinity
for all $(i,j)$, even the unobserved $(i,j)$ pairs. However, our model uses
$\xij^\tran\beta+\ai+\bj$ where both $\ai$ and $\bj$
are unbounded Gaussian random variables.
Now $\max_{1\le i\le R}\ai$ is asymptotically like $\sqrt{2\log(R)}\sigma_A$
and the $\bj$ satisfy a similar bound.
Thus we will assume that
\begin{align*}
\lim_{S\to\infty}\Pr\Bigl(\max_{1\le i\le R}\max_{1\le j\le C} |\xij^\tran\beta+\ai+\bj|
>\alpha\log(S)\Bigr) = 0
\end{align*}
holds for any $\alpha>0$.
Then for any $\psi>0$,
\begin{align}\label{eq:nobiggy}
\lim_{S\to\infty}\Pr\Bigl(
\min_{1\le i\le R}\min_{1\le j\le C} W_{ij}
< S^{-\psi}
\Bigr) = 0.
\end{align}

We need to bound the largest eigenvalue of $B_*B_*^\tran$
from equation~\eqref{eq:errorstobound}. In our context that matrix equals
\begin{align}\label{eq:ourbbtran}
\diag(\wid+\ssai)^{-1/2}
W
\diag(\wdj+\ssbi)^{-1}
W^\tran
\diag(\wid+\ssai)^{-1/2}.
\end{align}

\begin{proposition}\label{prop:lambda1}
Let $R$, $C$ and $\zij$ for $1\le i\le R$ and $1\le j\le C$
be sampled as in Section~\ref{sec:missing}
with $\rho,\kappa\in(0,1)$
and $\max(2\rho+\kappa,\rho+2\kappa)<2$.
Let $\lambda_1$ be the largest eigenvalue of
the matrix at~\eqref{eq:ourbbtran}.
Then if $\wij\le \bar\omega<\infty$
$$
\lim_{S\to\infty}\Pr
\biggl( \lambda_1 \le
\frac1{1+S^{\rho-1}/[(\pup+\epsilon)\bar\omega \ssa]}
\frac1{1+S^{\kappa-1}/[(\pup+\epsilon)\bar\omega \ssb]}
\biggr) = 1
$$
holds for any $\epsilon>0$.
\end{proposition}
\begin{proof}
By equations~\eqref{eq:gotnid} and~\eqref{eq:gotndj}
both
$$
\max_i\wid \le (\pup+\epsilon)\bar\omega S^{1-\rho}
\quad\text{and}\quad
\max_j\wdj \le (\pup+\epsilon)\bar\omega S^{1-\kappa}
$$
hold with probability tending to $1$ as $S\to\infty$
for any $\epsilon>0$.
Now
\begin{align*}
\lambda_1 &\le
\max_i\frac{\wid}{\wid+\ssai}
\times
\max_j\frac{\wdj}{\wdj+\ssbi}.
\end{align*}
Hence, for any $\epsilon >0$
$$
\lim_{S\to\infty}\Pr
\biggl( \lambda_1 \le
\frac1{1+S^{\rho-1}/[(\pup+\epsilon)\bar\omega \ssa]}
\frac1{1+S^{\kappa-1}/[(\pup+\epsilon)\bar\omega \ssb]}
\biggr) = 1.
$$
\end{proof}

\begin{theorem}\label{thm:eigvalso1}
Let $R=S^\rho$, $C=S^\kappa$ and $\zij$
follow the sampling model from Section~\ref{sec:missing}
for some $\pup<\infty$.
Assume that $\rho,\kappa\in(0,1)$
with $\max(2\rho+\kappa, \rho+2\kappa)<2$
and that $\Pr( \min_{ij}W_{ij}>S^{-\psi})\to1$
for all $\psi>0$
as described prior to equation~\eqref{eq:nobiggy}.
Let the matrix $B_*B_*^\tran$ in~\eqref{eq:ourbbtran}
have eigenvalues
$\lambda_1\ge\lambda_2\ge\cdots\ge\lambda_R\ge0$.
Then there exists $\alpha<1$ such that
for any  $\delta>0$
\begin{align}\label{eq:ratealpha}
\lim_{S\to\infty}\Pr\biggl(\,
\sum_{i=1}^R1\{\lambda_i>\delta\}
> R^\alpha \biggr)=0.
\end{align}
\end{theorem}
\begin{proof}
See Appendix~\ref{sec:proof:thm:eigvalso1}.
\end{proof}

\begin{theorem}\label{thm:trapproxtheorem}
Let $R=S^\rho$, $C=S^\kappa$ and $\zij$ for $1\le i\le R$ and $1\le j\le C$
be sampled as in Section~\ref{sec:missing}.
Assume that $\rho,\kappa\in(0,1)$
and that $\Pr( \min_{ij}W_{ij}>S^{-\psi})\to1$
for all $\psi>0$
as described prior to equation~\eqref{eq:nobiggy}.
Then our approximation error from~\eqref{eq:errorstobound} is
$$\err\equiv\tr\biggl( A^{-1}\sum_{k\ge1}(B_*B_*^\tran)^k\biggr)$$
and for any $\gamma>0$
$$\lim_{S\to\infty}\Pr( \err > \gamma R)=0.$$
\end{theorem}
\begin{proof}
See Appendix~\ref{sec:proof:thm:trapproxtheorem}.
\end{proof}

\subsection{Error with $\wh W$ versus $W$}\label{sec:wvswhat}
Our algorithm makes a trace approximation
to give Schall's algorithm a feasible cost.
We have proved that this approximation brings
a negligible error when the true
$\wij = \mu_{ij}(1-\mu_{ij})/\phi$
are used.
In practice the algorithm uses estimates
$\wh{W}_{ij}, \ssah, \ssbh$ and $\hat{\phi}$.
The principle difference
is that estimates $\wh{W}_{ij}$ might
be closer to zero than the true $\wij$.
We actually do not expect this to happen.
The process of estimating $\beta$ and $\bsa$
and $\bsb$ biases them towards the origin
and consequently away from very small weights.
We could in principle choose a small value $\psi$
and impose a minimum $\wh{W}_{ij}\ge N^{-\psi}$
in each stage of the algorithm but this has not been necessary.

We have also studied the algorithm
assuming that the true $\ssa, \ssb$ and $\phi$ are used,
while the algorithm runs with estimates $\ssah$, $\ssbh$
and $\hat\phi$.
The value of $\hat\phi$ is not consequential for Theorem~\ref{thm:eigvalso1} because
it scales all the weights $\wh W_{ij}=\hat\mu_{ij}(1-\hat\mu_{ij})/\hat\phi$
by the same factor leaving the weight ratios in the proof unchanged.
Smaller $\ssah$ and $\ssbh$
reduce the bound on $\lambda_1$ so they pose no difficulty.
Larger values of $\ssah$ and $\ssbh$
imply smaller ridge penalties on the
weighted least squares problems
and thereby increase the upper bound on the largest eigenvalue
$\lambda_1$ in Proposition~\ref{prop:lambda1}.
Very large values are a priori implausible for our target setting.
For instance, enormous $\ssa$ would tend to make
$\yij$ equal $0$ or $1$ depending almost entirely on the
row index $i$.
In practice one could impose a constraint like
$\max(\hat\sigma_A,\hat\sigma_B)\le 10$
but we have not had to do that.
For Theorem~\ref{thm:trapproxtheorem}, we only need $\ssah$ and
$\ssbh$ to be $o(S^{\delta})$ for any $\delta >0$.

\section{Clubbed backfitting}\label{sec:clubbing}

The iterative method in Section~\ref{subsec:modifiedschallapproach}
to solve~\eqref{eq:minboth} might suffer from
slow convergence because of
confounding of a factor with one of the variables $x_{ij\ell}$.
Let us consider the following optimization problem motivated by
the single factor problem~\eqref{eq:mina}:
\begin{equation}\label{eq:minobj}
\min_{\beta,\bsa}\|z-\cx\beta-\cza \bsa\|_{\bsw}^2 +\ssai\|\bsa\|^2.
\end{equation}
It may happen that a column of $\cx$ is exactly equal to the sum of
$k$ columns of $\cza$, e.g., columns $\ell_1, \ell_2,\ldots,\ell_k$. For example
if the rows in the data represent customers,
this column of $\cx$ could represent all male
customers older than 50. This linear dependence
has consequences for the solution $\bsa$ to (\ref{eq:minobj}).

\begin{lemma}\label{lem:sumzero}
  Suppose that column $q$ of $\cx$ in problem~\eqref{eq:minobj} equals the
  sum of $k\ge1$ distinct columns $\ci=\{i_1,i_2,\dots,i_k\}$ of $\cza$.
  If $\ssai>0$ then the solution for $\bsa$ satisfies
  $\sum_{\ell=1}^k a_{i_\ell}=0$.
\end{lemma}
\begin{proof}
For any $\bsa\in\real^R$, $\beta\in\real^p$  and $c\in\real$, define
$\bsa^{(c)}$ and $\beta^{(c)}$ via
\begin{align*}
a^{(c)}_{i}&=
\begin{cases}
a_{i}, &\text{if $i\in \ci$}\\
a_{i} - c, & \text{else}
\end{cases}
\qquad\text{and}\qquad
\beta_\ell^{(c)} =
\begin{cases}
\beta_\ell+c, &\text{if $\ell=q$}\\
\beta_\ell, & \text{else.}
\end{cases}
\end{align*}
If we evaluate the quadratic in~\eqref{eq:minobj} at $(\beta^{(c)},\bsa^{(c)})$
then the first term does not depend on $c$
and the second term has a unique minimum
at $c=(1/k)\sum_{i\in\ci}a_i$.
As a result $(\beta,\bsa)$ can only be the solution
if $c=(1/k)\sum_{\ell=1}^ka_{i_\ell}=0$.
\end{proof}

Lemma~\ref{lem:sumzero} imposes a constraint
on $\bsa$ for every column of $\cx$ that equals
a sum of columns of $\cza$.
Backfitting will eventually converge to a solution satisfying those
constraints, but it could take a long time to get there. We can speed
things up by enforcing any known constraints.
When as usual, $\cx$ includes a column of ones for an intercept
we get the constraint $\sum_{i=1}^Ra_i=0$,
as a special case.

With two factors we can also have such aliasing with sums of columns of
$\czb$ as well, with particular columns of $\cx$. For example a column
of $\cx$ might represent a particular category of clothing,
corresponding to a number of levels of the column factor.
Furthermore there may be  other non-trivial constraints that are implied by $\cx$.
As we saw in equation~\eqref{eq:minboth}, we want to obtain $\displaystyle\arg\min_{\beta,\bsa,\bsb} \pl(\beta,\bsa,\bsb)$ for
the penalized least squares problem
\begin{equation}\label{eq:minab}
\pl(\beta,\bsa,\bsb) = \sum_{i,j} \zij\wh{W}_{ij} \left( z_{ij}  -\xij^\tran\beta - a_{i} - b_{j}\right)^2
+\ssiha\Vert\bsa\Vert^2 + \ssihb\Vert\bsb\Vert^2.
\end{equation}
We use the following iterative ``clubbing'' strategy to enforce all
of these constraints automatically.
Given $(\beta^{(k)},\bsa^{(k)},\bsb^{(k)})$ we first
optimize $\pl(\cdot)$ over $\beta$ and $\bsa$
with $\bsb=\bsb^{(k)}$ fixed to get
$\bsa^{(k+1)}$ and an intermediate quantity $\beta^{(k+\frac12)}$.
Then we optimize $\pl(\cdot)$ over $\beta$ and $\bsb$
with $\bsa=\bsa^{(k+1)}$ fixed to get
$\bsb^{(k+1)}$ and the next iterate $\beta^{(k+1)}$ of the regression vector.

It is convenient to describe the updates in terms
of fitted quantities in $\real^N$.
The needed parts $\beta^{(k)}$, $\bsa^{(k)}$ and $\bsb^{(k)}$
are easily obtained in the process.
For the first part of the iteration we set
\begin{align}\label{eq:factora}
\begin{split}
\cx\beta^{(k+\frac12)} &
= (\cx^\tran \widehat{\cw}(I_N-\cS_A)\cx)^{-1}\cx^\tran \widehat{\cw}(I_N-\cS_A)(z-\czb
\bsb^{(k)}),\quad\text{and}\\
\cza\bsa^{(k+1)} &= \cS_A(z-\cx{\beta}^{(k+\frac12)}-\czb{\bsb^{(k)}})
\end{split}
\end{align}
for a smoother matrix
$\cS_A = \cza(\cza^\tran \widehat{\cw} \cza +
\ssiha I_R)^{-1}\cza^\tran \widehat{\cw}$
that simply computes weighted
group means.
This gets applied to each column of $\cx$ in the
first equation, and to the residual in the second.
The matrix inversion in the formula is handled by solving a $p\times p$
system of equations which adds a cost that is of constant
order in $N$.

The equations~\eqref{eq:factora}
solve the following minimization problem,
\begin{equation}\label{mina}
\min_{\beta,\bsa}\sum_{i,j} \zij\wh{W}_{ij} \bigl( z_{ij} -{b_j^{(k)}} -\xij^\tran\beta-\ai\bigr)^2
+\ssiha\Vert\bsa\Vert^2.
\end{equation}
If we absorb $b_j^{(k)}$ into $z_{ij}$ we see that~\eqref{mina} has the
form~\eqref{eq:minobj} and so by Lemma~\ref{lem:sumzero}
the solution $(\beta^{(k+\frac12)},\bsa^{(k+1)})$ satisfies any constraint that is intrinsic to the design.

To complete the iteration we fix $\bsa = \bsa^{(k+1)}$ and
optimize over $\beta$ and $\bsb$, via
\begin{align*}
\begin{split}
\cx\beta^{(k+1)} &= \cx(\cx^\tran \widehat{\cw}(I_N-\cS_B)\cx)^{-1}\cx^\tran \widehat{\cw}(I_N-\cS_B)(z-\cza{\bsa^{(k+1)}}),\quad\text{and}\\
\czb\bsb^{(k+1)} &= \cS_B(z-\cx{\beta^{(k+1)}}-\cza{\bsa^{(k+1)}})
\end{split}
\end{align*}
for a smoother matrix $\cS_B = \czb(\czb^\tran\wh\cw\czb+\ssihb I_C)^{-1}\czb^\tran\wh\cw$.
Once again Lemma~\ref{lem:sumzero} applies to the solution.

\begin{lemma}\label{lemma:clubbingconvergence}
If $\cx$ is of full rank, and $\max(\ssah,\ssbh)<\infty$ and the weights $\wh{W}_{ij}$ are positive, then the iterative algorithm converges to a global minimum of the equation~\eqref{eq:minab}.
\end{lemma}
\begin{proof}
See Appendix~\ref{sec:proof:lem:clubbingconvergence}.
\end{proof}

We can compute these single factor operators in $O(N)$ computation and use them for each iteration of the inner (solving the weighted least squares problem for a fixed weight vector) backfitting loop rather than computing it every time we run this iteration. The iterative algorithm stops when the relative change in $\zeta \equiv \cx\beta+\cza\bsa+\czb\bsb$ is below a certain threshold. At convergence of the clubbed backfitting, we obtain $\hat{\beta}, \hat{\bsa}, \hat{\bsb}$ for a particular set of weights.

Next we update the weights using the new set of parameters and solve a new optimization problem until convergence. We stop when the relative change in the fitted values $\zeta$ obtained with a different set of weights is negligible.
Not only does this clubbed variant of block-coordinate descent automatically satisfy the implicit constraints,
but we expect (and indeed observe) faster convergence than other variants that do not enforce the constraints.
Here is a concise description of the modified Schall algorithm with clubbed backfitting:
\begin{compactenum}[\quad 1)]
\item Stage $\ell$ of the algorithm provides $\hat\beta^{(\ell)}$, $\hat{\bsa}^{(\ell)}$, $\hat{\bsb}^{(\ell)}$ using the weights obtained in the $(\ell-1)$'st iteration and clubbing method described earlier. Obtain $\hat{\zeta}^{(\ell)} = \cx\hat{\beta}^{(\ell)} + \cza \hat{\bsa}^{(\ell)} + \czb \hat\bsb^{(\ell)}.$
\item Obtain $\hat\sigma_A^{2(\ell)}$,  $\hat\sigma_B^{2(\ell)}$, and $\hat\phi^{(\ell)}$ using the trace approximation.
\item Recompute the weights with new parameters.
\end{compactenum}
\smallskip

We iterate until
$$
\frac{\Vert \hat{\zeta}^{(\ell)}-\hat{\zeta}^{(\ell-1)}\Vert^2}{\Vert\hat{\zeta}^{(\ell-1)}\Vert^2}
< \epsilon
$$
and then deliver $\hat\beta^{(\infty)} =\hat\beta^{(\ell)}$, $\hat\bsa^{(\infty)} =\hat\bsa^{(\ell)}$,
$\hat \bsb^{(\infty)}=\hat\bsb^{(\ell)}$, $\hat\sigma_A^{2(\infty)}=\hat\sigma_A^{2 (\ell)}$, $\hat\sigma_B^{2(\infty)}=\hat\sigma_B^{2 (\ell)}$
and $\hat\phi^{(\infty)} = \hat\phi^{(\ell)}$. Our generalized linear mixed model coefficient estimate is then
$\hat\beta_{\glmm} = \hat\beta^{(\infty)}$.
Once we obtain the weights and parameter estimates at convergence
we can compute $\wh\cov(\hat\beta_{\glmm})$ by  backfitting on each
column of $\cx$ using
$\hat\sigma_A^{2(\infty)}$, $\hat\sigma_B^{2(\infty)}$, $\wh{\cw}^{(\infty)}$ at~\eqref{eq:covbetahat2factor}. As
a consequence of Lemma~\ref{lem:sumzero}, if $\cx$ contains the intercept the penalized least squares
problem in~\eqref{eq:minboth} is equivalent to solving a constrained
penalized least squares problem with constraint $\sum_{i=1}^{R} a_{i} = 0$
and $\sum_{j=1}^{C} b_{j} = 0$. We can make the algorithm more efficient
if we use centered operators (See \cite{ghos:hast:owen:2021})
$\tilde \cS_A^{(k)}$ and $\tilde \cS_B^{(k)}$ in ~\eqref{eq:backfittwoblocks}
 instead of $\cS_A^{(k)}$ and $\cS_B^{(k)}.$

Our numerical results in the Section~\ref{sec:empiricalresults} use $\epsilon =10^{-8}$.
Open-source R code at
\url{https://github.com/G28Sw/backfitting_binary_regression}
does these computations.

\section{Timing and accuracy comparisons}\label{sec:empiricalresults}

In this section we simulate the data from a crossed random effects
model with a binary response.
A simulated setting with known $\beta$ lets us compare
accuracy of the methods.  We also can verify linear costs for backfitting
and superlinear costs for a state of the art algorithm,
glmer from \cite{lme4}.
We assume that the observation pattern
follows the probability model from Section~\ref{sec:missing}.
We look at timings and we also consider accuracy.
These experiments were carried out in R on a computer
with the macOS operating system,  16 GB of memory and an Intel i7 processor.

\subsection{Timings}
To study how the computation time varies with $N$
we generated data over a range of sample sizes
using $\rho = \kappa = 0.56$.
Our predictors were $\xij\simiid\dnorm(0,\Sigma)$
in seven dimensions, plus an intercept, making $p=8$.
The covariance matrix had $\Sigma_{k\ell} = \gamma^{|k-\ell|}$
with autoregression parameter $\gamma = 1/2$.
We took $\beta = \boldsymbol{0}$, $\sigma_{A}=0.8$ and $\sigma_{B} = 0.4$.

We included our backfit iteratively reweighted least squares
algorithm, a naive logistic regression ignoring random effects
and two versions of the glmer R code.
One version of glmer is the default using $1$ Gaussian quadrature point.
The other uses no such points and is obtained by calling glmer with the
option {\tt nAGQ = 0}.
Not surprisingly, we will see that this version of glmer is faster than the default, but less accurate.

Theory suggests the computation time for glmer should be of the order
$N^{\max(3\rho,3\kappa)}$. So we expect the order of computation to be
$N^{1.68}$ in this case, whereas Schall backfitting with the trace approximation
should have a cost that is linear in $N$.

\begin{figure}
\centering
  \includegraphics[width=.48\textwidth]{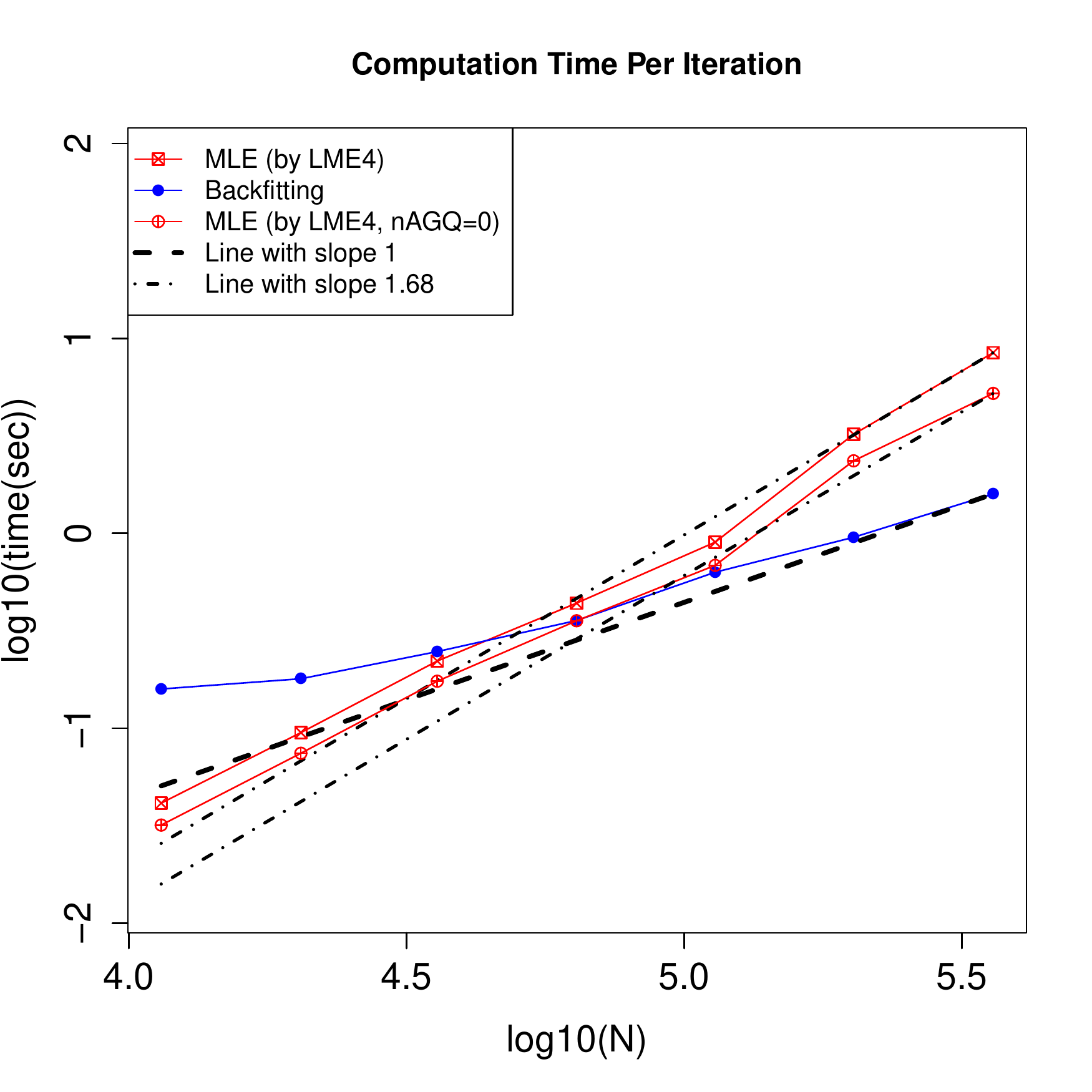}
\includegraphics[width=.48\textwidth]{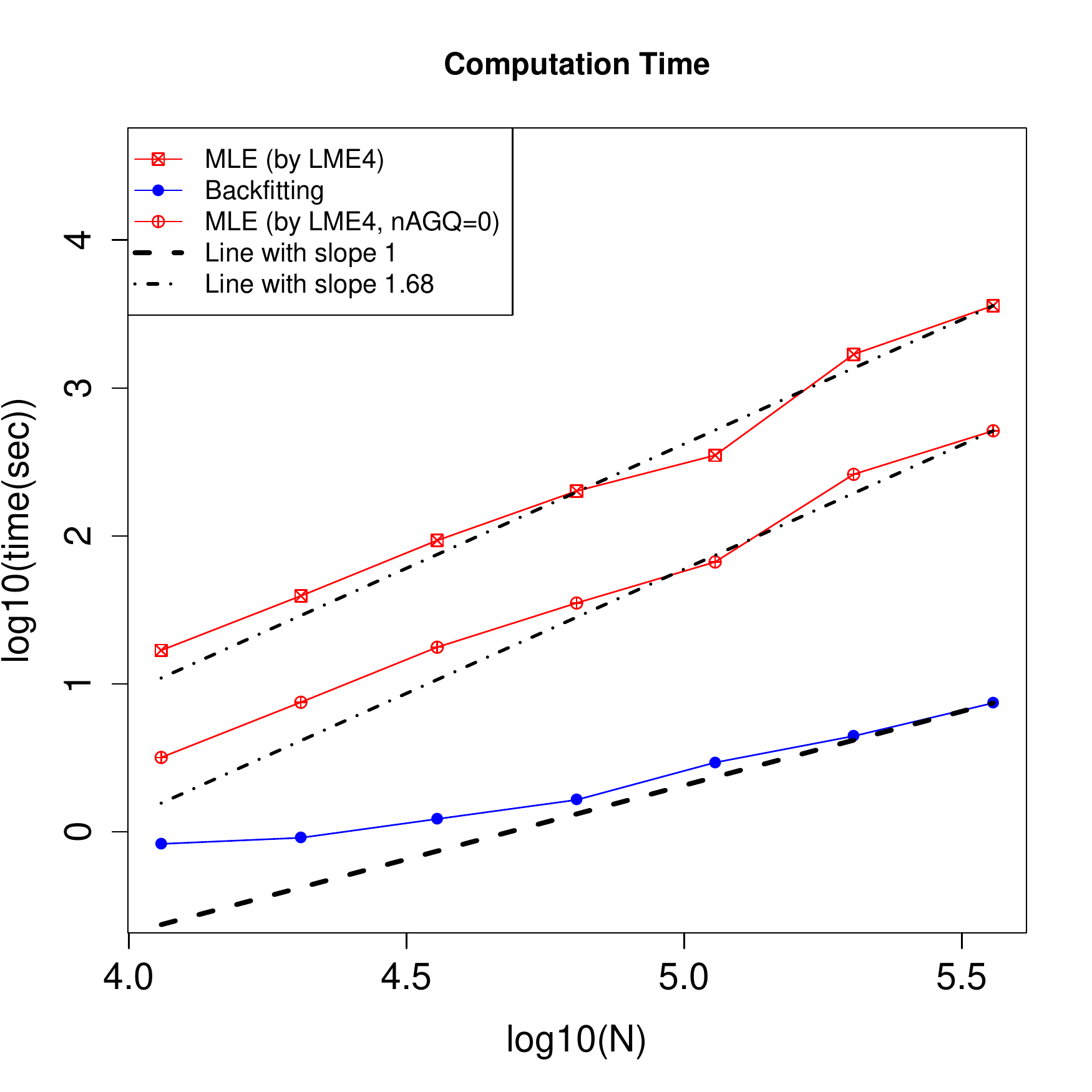}

\caption{\label{fig:comptimes1singlelinenagq}
Time for one iteration and total time versus $N$ at $(\rho,\kappa) = (0.56,0.56)$.
The cost for glmer is roughly $O(N^{\max(3\rho,3\kappa)})$.  Costs for backfitting
is $O(N)$.
}
\end{figure}

Figure~\ref{fig:comptimes1singlelinenagq} shows our timings.
The left panel shows the time required for a single iteration.
We see that for the largest values of $N$ in our simulation,
the cost of one single iteration of glmer starts to follow
a trend close to $N^{1.68}$. As expected the default algorithm is
slower. It appears to be slower by a constant factor.
The cost of backfitting appears to grow more slowly
than $O(N)$ over the given range but we believe this
is because of startup costs and we see that by the end
of the range of $N$ values that cost is growing nearly
proportionally to $N$ as expected.

The cost of an algorithm depends on not just the cost
per iteration but also on the number of iterations.
The second panel of
Figure~\ref{fig:comptimes1singlelinenagq} shows times
to convergence. At the largest values of $N$, the timings follow the
same asymptotic rates, approximately $O(N^{1.68})$
and $O(N)$, as we saw for individual iterations.
The costs tend to approach these asymptotes from above.
That could be startup costs or it might arise because
the number of iterations tends to decrease with~$N$.

\subsection{Accuracy}
It is also important to compare the accuracy of these
algorithms.
We also compute the $\hat{\beta}_{\mathrm{GLMER}}$, $\hat{\beta}_{\mbox{Backfitting Schall}}$ and $\hat{\beta}_{\mathrm{Logistic}}$ and compute the MSE. We used two different set of $\beta$:
\begin{compactenum}[\quad a)]
\item $\beta_{\ell} = 0$ for $\ell=1,\ldots,7$ and $\beta_{0} = -2$, and
\item $\beta_{\ell} = -2 + 0.5\times\ell$ for $\ell=1,\dots,7$ and $\beta_{0} = -2$.
\end{compactenum}
The choice of negative $\beta_{0}$ was made to simulate a setting
where $\yij=1$ is a somewhat rare event.
Some consequences for naive logistic regression
of nonzero $\beta$ are described in Section~\ref{sec:logistic}.
We ran 100 replicates for each value of $S$ and then computed the MSE of
the estimators $\hat\beta$, $\hat\sigma_{A}$ and $\hat\sigma_{B}$.
For accuracy we chose $S$ to get
$\log_{10}(N)$ in the range from $3$ to $4.5$ for
glmer algorithms.  Because naive logistic regression and
backfitting are $O(N)$ we studied them over a range
up to $\log_{10}(N)\approx 5.5$.

\begin{figure}
\centering
  \includegraphics[width=.48\textwidth]{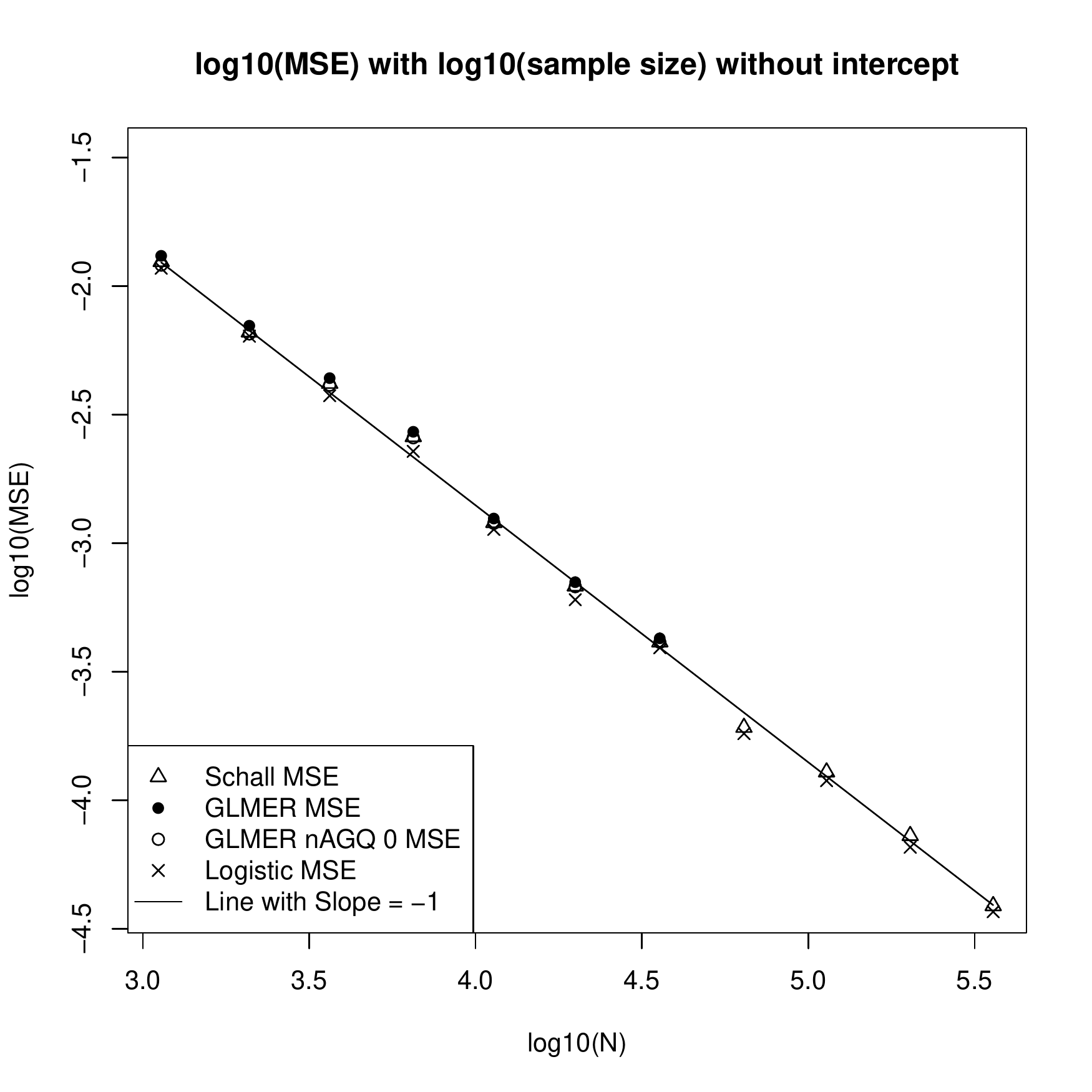}
\includegraphics[width=.48\textwidth]{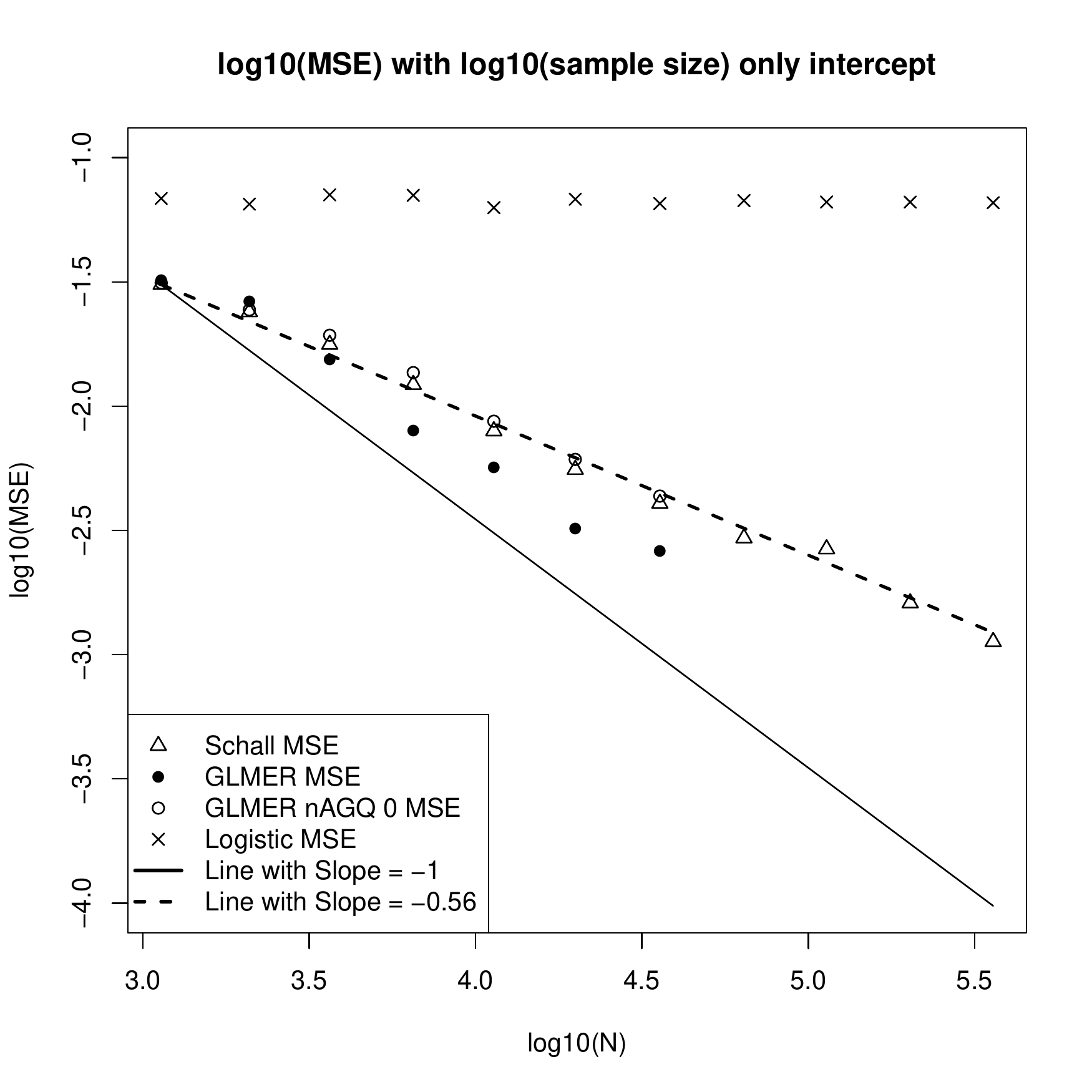}

\caption{\label{fig:mse_zero_beta}
MSE of $\hat{\beta}$ vs $N$ at $(\rho,\kappa) = (0.56,0.56)$ for the choice of $\beta$ in a).
MSE for glmer, backfitting and Schall
seems to scale at $O(1/N)$ for non intercept regression coefficients. MSE for intercept seems to scale at a slower rate compared to $O(1/N)$, whereas logistic shows no improvement with increasing problem size.
}
\end{figure}

Figure~\ref{fig:mse_zero_beta} handles case a) above with all $\beta_\ell=0$
except for the intercept.
The left panel shows the MSE for all $\beta_\ell$ except
the intercept.  The right panel shows the intercept.
The intercept error in naive logistic regression
does not decrease with $N$ although the error
for the other coefficients does.
The RMSE for non-intercept components is close to
a reference line parallel to $O(1/N)$ and this holds
for backfitting, naive logistic regression and both
glmer algorithms. The RMSE for the intercept
follows close to $O(1/\sqrt{N})$ for backfitting
and glmer with {\tt nAGQ=0}.
The RMSE for default glmer appears to get
a rate in between $O(1/\sqrt{N})$ and $O(1/N)$.
We had expected the intercept coefficient
to show slow convergence based on its
partial confounding with the random effects
described in the introduction.
If the true MSE rate for the intercept really
is $o(N^{-1/2})$ for default glmer, then we are unable
to explain that.

\begin{figure}
\centering
  \includegraphics[width=.48\textwidth]{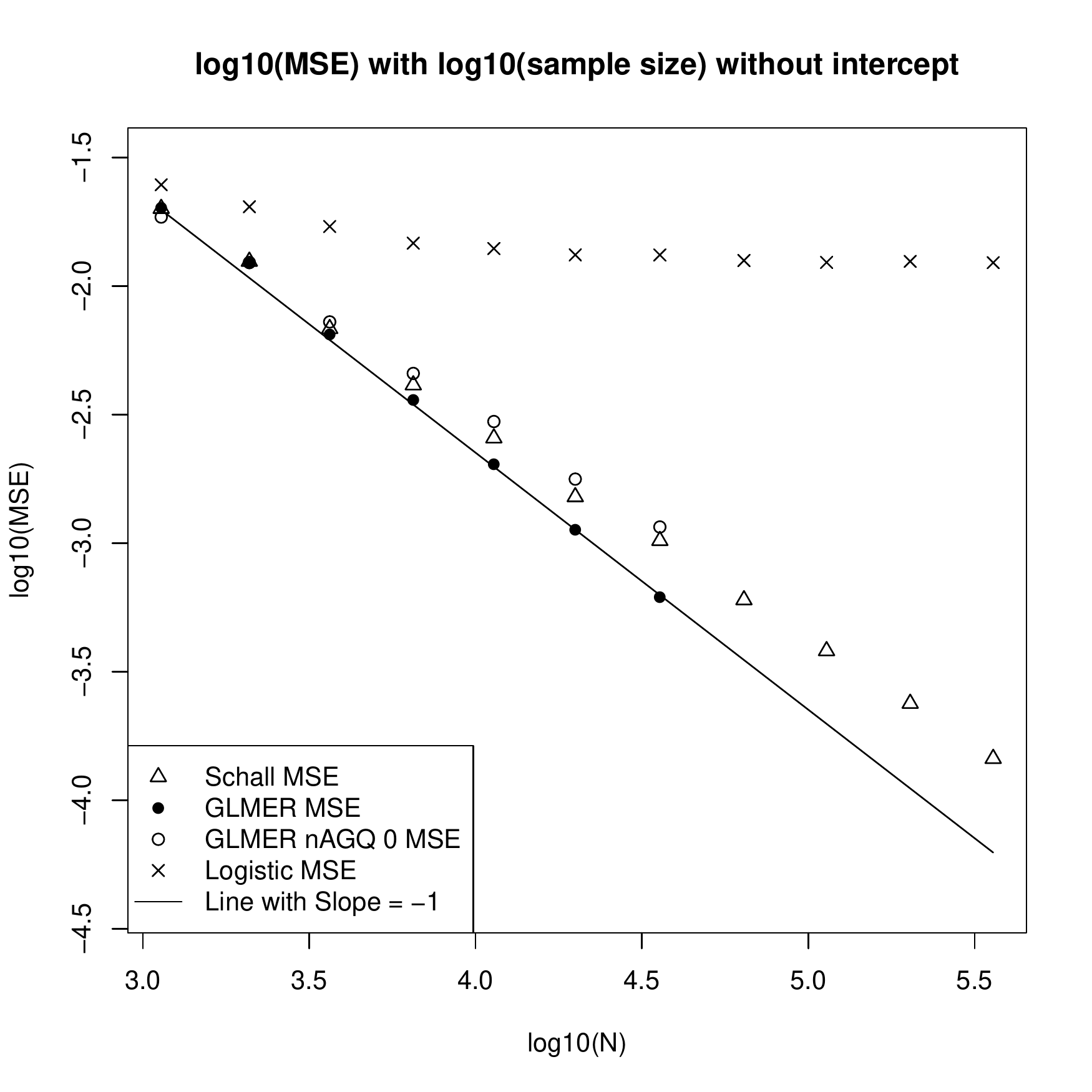}
\includegraphics[width=.48\textwidth]{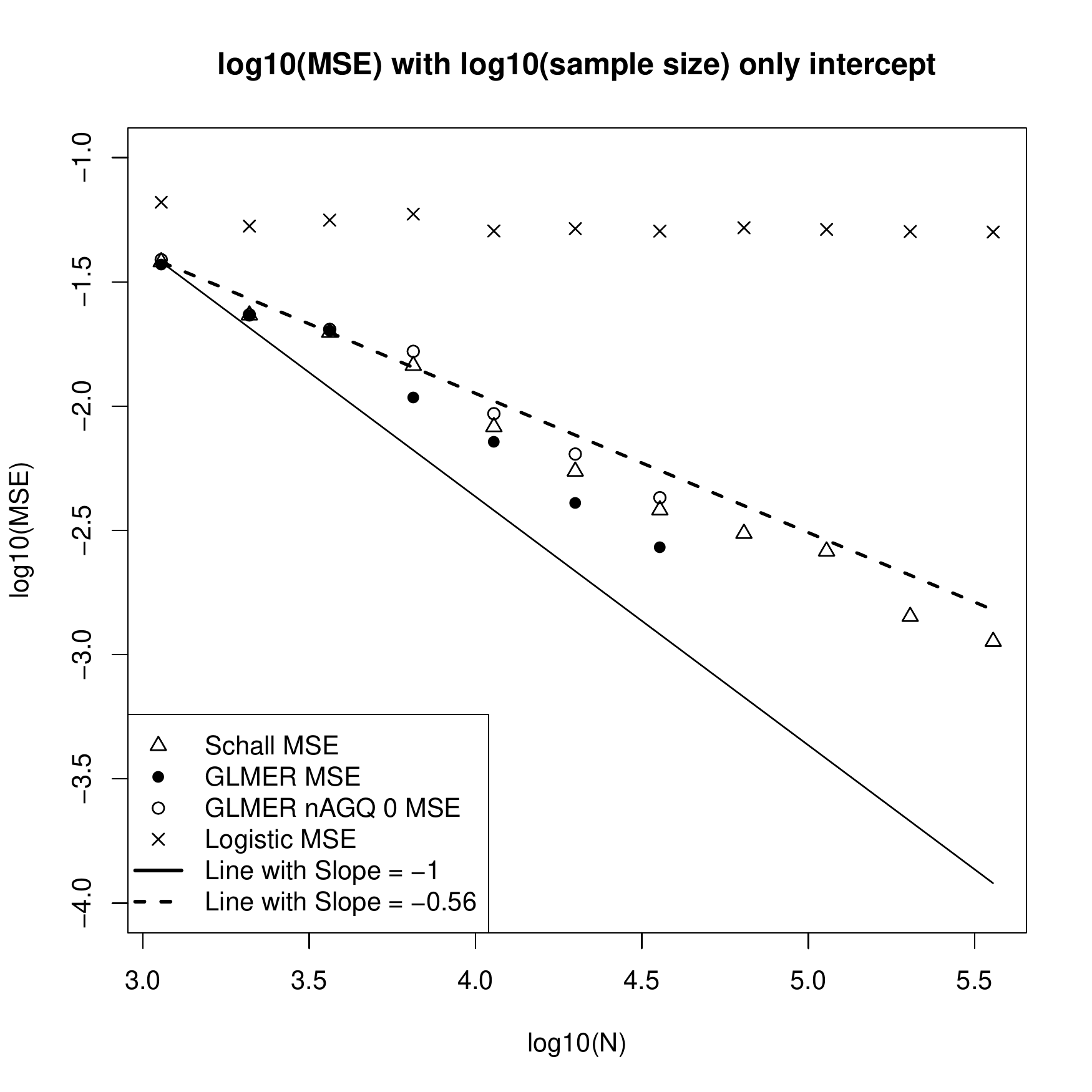}

\caption{\label{fig:mse_non_zero_beta}
MSE of $\hat{\beta}$ vs $N$ at $(\rho,\kappa) = (0.56,0.56)$ for the choice of $\beta$ in b).
MSE for glmer and backfitting
seems to scale at $\approx O(1/N)$ whereas logistic shows no improvement with increasing problem size. MSE for intercept seems to scale at a slower rate compared to $O(1/N).$
}
\end{figure}

Figure~\ref{fig:mse_non_zero_beta}
shows case b) with nonzero $\beta_\ell$.
Plain logistic regression appears to be inconsistent
for both intercept and non-intercept parameters.
Once again backfitting is less accurate than
the default glmer but more accurate than the
other glmer choice. The default glmer has
an RMSE close to $O(N^{-1})$ for non-intercept
parameters. Backfitting and the faster glmer
show evidence of having a worse than $O(N^{-1})$
convergence rate for non-intercept terms.

\subsection{About naive logistic regression}\label{sec:logistic}

This section provides an explanation
for some of the biases we see for naive
logistic regression.  This analysis preceded
the simulations and influenced our
choice of simulation settings.

Suppose we have a mixed effects model
with one random effect (no crossing).
The responses $\yij$ satisfy
$\Pr(\yij=1\giv \xij, a_i ) =
\pi(\xij^\tran\beta+a_i)
$
for $i=1,\dots,R$. Since the columns are
not linked we let $j=1,\dots,\nid$
The random effects are
$a_i\simiid\dnorm(0,\sigma^2)$.
The $\xij$ are fixed either by design or by running
a conditional analysis.
We will suppose at first that $\sigma$ is `small'.

If an analyst ignores the random effects and uses likelihood
$$
\tilde L = \prod_{i=1}^R\prod_{j=1}^{\nid}
\pi(\xij^\tran\beta)^{\yij}
(1-\pi(\xij^\tran\beta))^{1-\yij}
$$
then their log likelihood function is
$$
\tilde \ell
=\sum_{i=1}^R\sum_{j=1}^{\nid}
\yij \xij^\tran\beta-\log(1+\exp(\xij^\tran\beta)).
$$
Their naive score function is
$$
\frac{\partial \tilde\ell}{\partial\beta}
=\sum_{i=1}^R\sum_{j=1}^{\nid}
\bigl(\yij-\pi(\xij^\tran\beta) \bigr)\xij.
$$
Their MLE will be consistent if and only if
the expected score (under the random effects model)
is zero at the true $\beta$.
Randomness enters that expectation via $\yij$
which includes randomness from $a_i$.

Now
\begin{align*}
\e\biggl(\frac{\partial \tilde\ell}{\partial\beta}\biggr)
=\sum_{i=1}^R\sum_{j=1}^{\nid}
\biggl(\,
\int_{-\infty}^\infty
\frac{\exp(-a_i^2/(2\sigma^2))}{\sqrt{2\pi}\sigma}
\pi(\xij^\tran\beta+a_i)\rd a_i-\pi(\xij^\tran\beta) \biggr)\xij.
\end{align*}
For small $\sigma$, we anticipate that small $a$ will dominate,
so we make a Taylor approximation
\begin{align*}
\pi(\xij^\tran\beta+a_i)
\approx \pi(\xij^\tran\beta)
+a_i\pi'(\xij^\tran\beta)
+\frac12a_i^2\pi''(\xij^\tran\beta)
\end{align*}
and using this approximation
\begin{align}\label{eq:needortho}
\e\biggl(\frac{\partial \tilde\ell}{\partial\beta}\biggr)
\approx \frac{\sigma^2}2\sum_{i=1}^R\sum_{j=1}^{\nid}
\pi''(\xij^\tran\beta)\xij.
\end{align}

In light of equation~\eqref{eq:needortho}, the analyst
needs each variable $x_{ij\ell}$ to be nearly
orthogonal to $\pi''(\xij^\tran\beta)$ with
the true unknown $\beta$.
This is virtually impossible to arrange.

In one special case we could have
$\beta$ equal to zero apart from
a negative intercept component.
Then $\pi''(\xij^\tran\beta)$ is a positive
constant and for any centered variables $x_{ij\ell}$
the $\ell$'th component of~\eqref{eq:needortho}
will vanish.  Of course the intercept variable
is all ones and cannot be centered and it will
therefore have a non vanishing score component
and hence a bias.

\section{Stitch Fix data example}\label{sec:stitchfix}

Stitch Fix is an online personal styling service.
One of their business models involves sending
customers a sample of clothing items. The customer may
keep and purchase any of those items and
return the others.
They have provided us
with some of their client ratings data.
That data was anonymized, void of personally identifying
information, and as a sample it does not reflect their total
numbers of clients or items at the time they provided it.
It is also from 2015. While it does not describe their
current business, it is a valuable data set for illustrative purposes.
The binary response of interest was whether
the customer thought an item was a top rated fit.
There were $N=5{,}000{,}000$ ratings from $744{,}482$ clients on $3{,}547$ items.
We want to treat both clients and items as random effects.
The data are not dominated by a single row or column because
$\max_i\nid/N\doteq 1.24\times 10^{-5}$ and $\max_j\ndj/N\doteq 0.0276$.
The data are sparse because $N/(RC)\doteq 0.0018$.

One of the predictors was the primary material of which
an item was made with $20$ levels such as `linen' or `wool'.
The material is a property of the item or to put it another
way, the item factor is nested within the levels of the material predictor.
The other predictors for this response were
`client dress size', `client chest size', `client fit profile' (later abbreviated as cfp),
`item fit profile' (later abbreviated as ifp), `client edgy' and `client boho'.
For instance the fit profiles had these
levels: `Fitted',  `Loose',  `Oversize',  `Straight' and `Tight'
as well as `Missing'.
Some clients indicated that they like `edgy' styles
and some that they like `boho' styles.

In a business setting one would fit and compare
a wide variety of different binary regression models
in order to understand the data.  Our purpose here
is to understand large scale generalized linear mixed
effects models and so we choose just one model
for illustration.
That model has
\begin{align}
\logit(\Pr(\yij=1\giv \ai,\bj))
& =  \beta_0+\beta_1\mathrm{client\ fit\ profile}_{i}+\beta_2\mathbb{I}\{\mathrm{client\  edgy}\}_i \notag \\
&\phe + \beta_3\mathbb{I}\{\mathrm{client\ boho}\}_i
  +\beta_4\mathrm{client\ chest\ size}_{i} \notag \\
  &\phe +\beta_5\mathrm{client\  dress\ size}_i +\beta_6\mathrm{material}_{j} \notag \\
  &\phe + \beta_7\mathrm{item\ fit\ profile}_{j}+a_i+b_j+e_{ij}. \notag
\end{align}

The categorical variables were incorporated using
a one-hot encoding with a binary indicator variable
for each level other than the most common level.
The model has $p = 34$ parameters.  Our backfitting algorithm took 14 iterations to convergence with a tolerance of $10^{-8}.$

Let $\hat\beta_\LR$ and $\hat\beta_\glmm$ be the logistic
regression and GLMM
estimates of $\beta$ obtained by clubbed backfitting, respectively. We can compute their
corresponding variance estimates
$\wh\cov_\LR(\hat\beta_\LR)$ and $\wh\cov_\glmm(\hat\beta_\glmm)$.
We can also find
$\wh\cov_\glmm(\hat\beta_\LR)$, the variance under our GLMM model of the
$\hat\beta_\LR$. The estimated coefficients $\hat\beta_\LR$, $\hat\beta_\glmm$ and their standard errors are presented in
Appendix~\ref{sec:stitchfixresults}.

\begin{figure}
\centering
\includegraphics[width=.4\hsize]{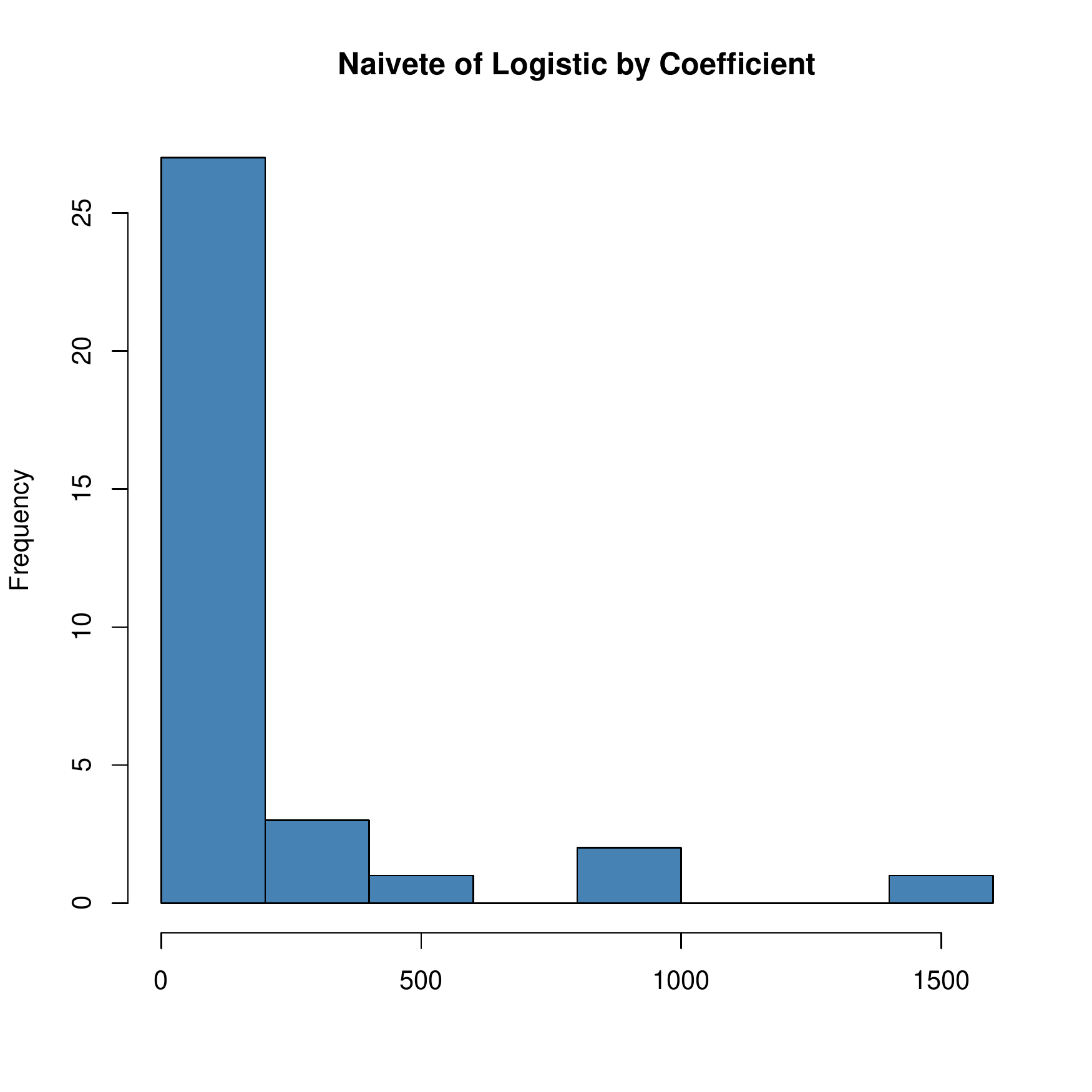}
\includegraphics[width=.4\hsize]{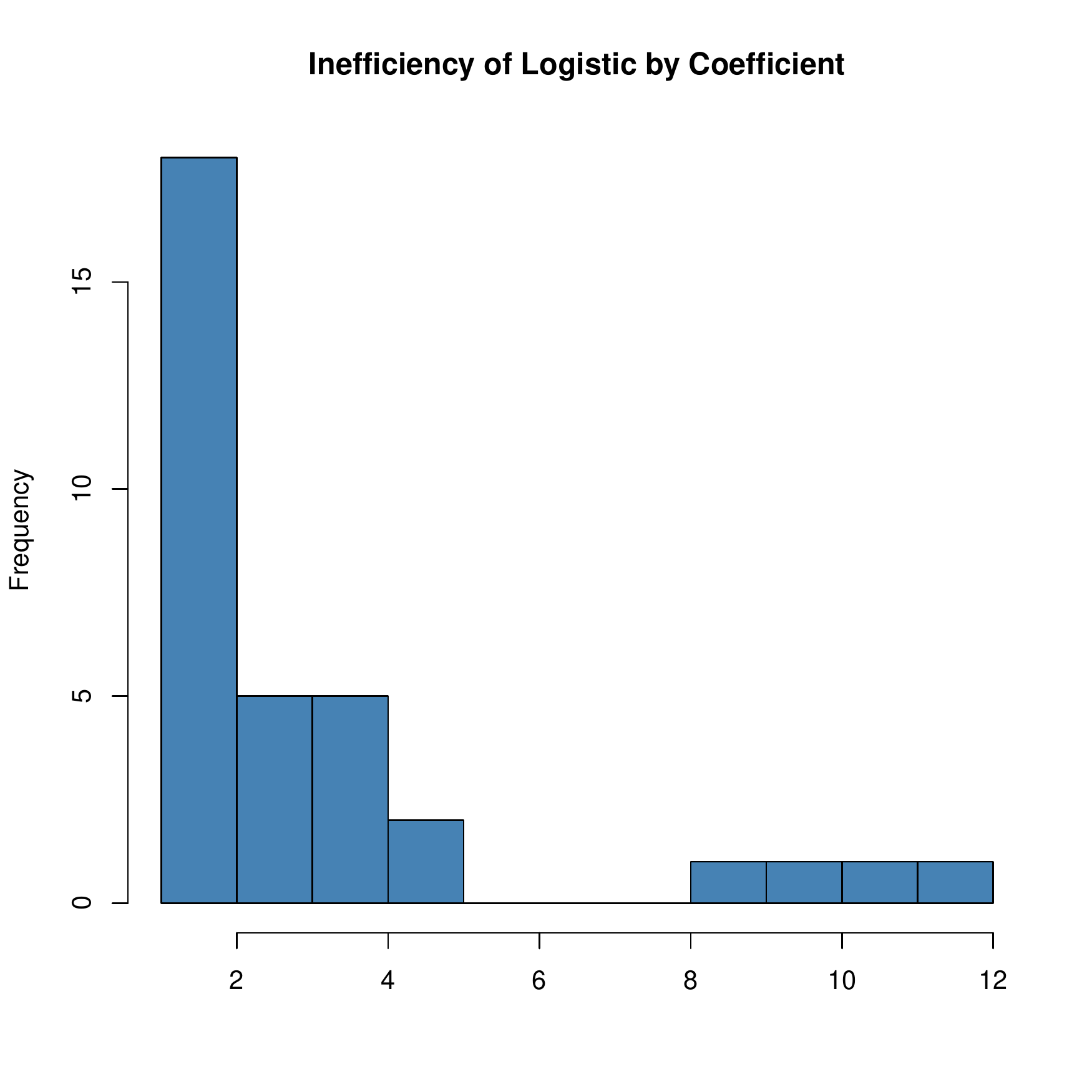}
\caption{\label{fig:logisticnaive}
The left panel is a histogram of
naivete of logistic regression quantified by
$\wh\cov_{\glmm}(\hat\beta_{\LR,\ell})/\wh\cov_{\LR}(\hat\beta_{\LR,\ell})$
for coefficients $\beta_\ell$ in the Stitch Fix data.
The right panel is the inefficiency
$\wh\cov_{\glmm}(\hat\beta_{\LR,\ell})/\wh\cov_{\glmm}(\hat\beta_{\glmm,\ell})$.
}
\end{figure}

We can quantify the naivete of logistic regression, coefficient by coefficient, via the ratio
$\wh\cov_{\glmm}(\hat\beta_{\LR,j})/\wh\cov_{\LR}(\hat\beta_{\LR,j})$.
The left panel of Figure~\ref{fig:logisticnaive} plots these values.
They range from $ 2.72$ to $1467.05$ and can be interpreted as
factors by which logistic regression naively overestimates its sample size.
The largest and second largest ratios are for material indicators
corresponding to `Modal' and `Rayon', respectively.
Not only is logistic regression estimating $\beta$ with significant
bias, it greatly underestimates its sampling uncertainty.
We can also identify the linear combination of $\hat\beta_\LR$
for which $\LR$ is most naive. We maximize
the ratio
$$\bsx^\tran\wh\cov_{\glmm}(\hat\beta_{\LR})\bsx/\bsx^\tran\wh\cov_{\LR}(\hat\beta_{\LR})\bsx$$
over $\bsx\ne0$.
The resulting maximal ratio is the largest eigenvalue of
$$\wh\cov_{\LR}(\hat\beta_{\LR}) ^{-1}
\wh\cov_{\glmm}(\hat\beta_{\LR})$$
and it is about $1507$ for the Stitch Fix data.

We can also quantify the inefficiency of logistic regression, coefficient by coefficient, via the ratio
$\wh\cov_{\glmm}(\hat\beta_{\LR,\ell})/\wh\cov_{\glmm}(\hat\beta_{\glmm,\ell})$.
The right panel in Figure~\ref{fig:logisticnaive} plots these values.
They range from just over $1$
to $11.42$ and can be interpreted as factors by which using
logistic regression reduces the effective sample size.  The two largest inefficiencies
corresponds to item material `Rayon' and `Modal' respectively.
The most inefficient linear combination of $\hat\beta$ reaches a
variance ratio of $15.50$.

\begin{figure}[t]
\centering
\includegraphics[width=.8\hsize]{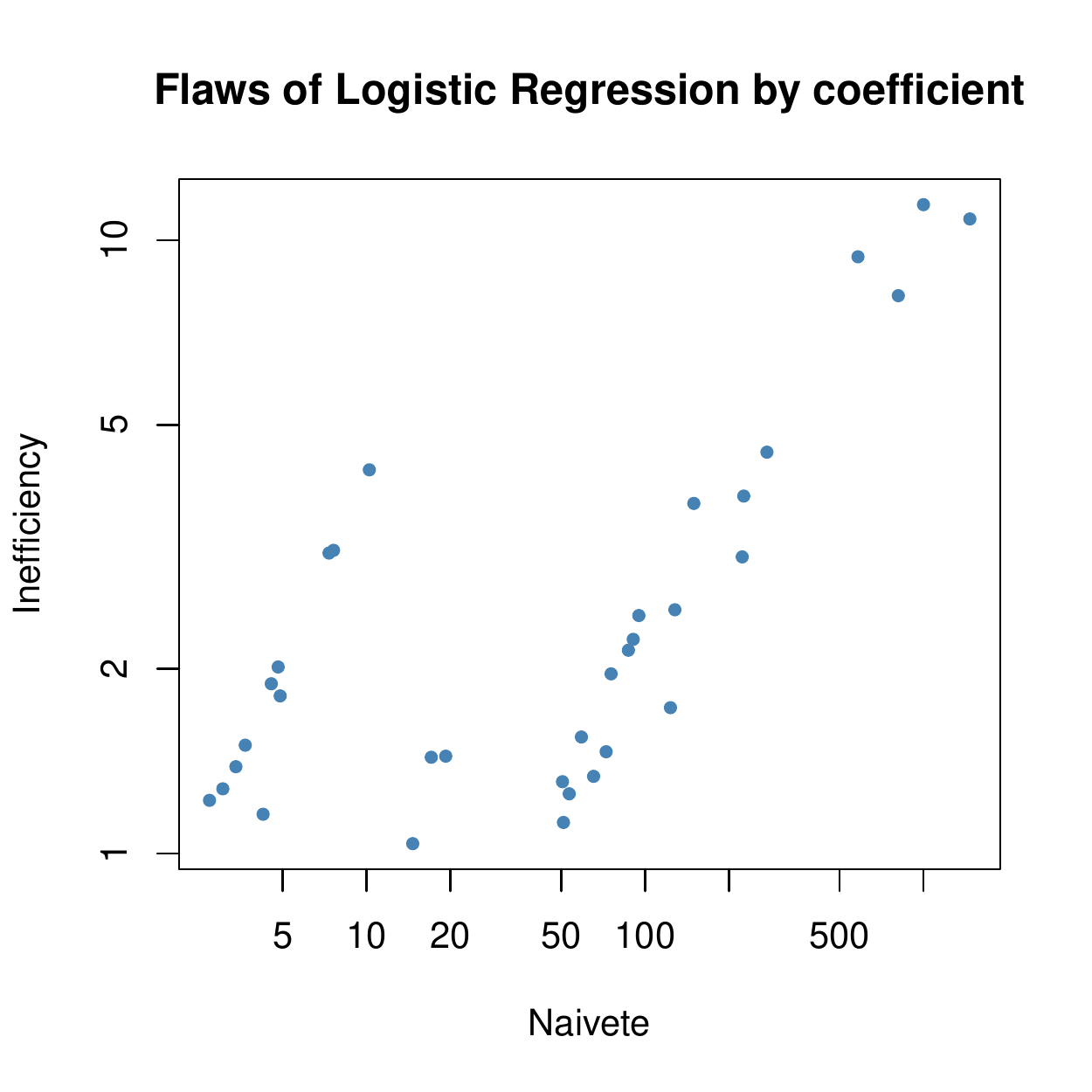}
\caption{\label{fig:naivevsinefficient}
Inefficiency vs naivete for logistic regression coefficients in the Stitch Fix data.
}
\end{figure}

Figure~\ref{fig:naivevsinefficient} plots inefficiency
versus naivete
for the $34$ coefficients in our logistic regression model.
The very worst coefficients by one measure tend to be
worst by the other as well.

\section{Discussion}\label{sec:discussion}

The most critical inference problems are at the margins
where we can just barely resolve real effects from noise.
Ignoring a crossed random effects correlation structure
can be very naive, underestimating the true sampling variance
of a parameter by several hundred-fold, leading to very unreliable
findings lacking reproducibility and even internal validity
much less external validity.

The crossed random effects setting makes
many of our standard methods
going back to \cite{SCM09} and \cite{H53}
computationally prohibitive.  A cost that grows faster than $N^{3/2}$
is not possible for somebody with big data.

There has been recent progress in speeding
up these computations including frequentist approaches for
least squares problems with crossed effects in
\cite{gao:thesis},
\cite{crelin},
and \cite{ghos:hast:owen:2021}
as well as Gibbs sampling in \cite{papa:robe:zane:2020}.
Our contribution is to extend the backfitting approach
to logistic regression.
A critical step was our clubbing
of $\hat{\beta}$ with one random effect at a time.
This can be viewed as a frequentist counterpart to the collapsed Gibbs
sampler in \cite{papa:robe:zane:2020}.
There remains a theoretical gap between what the
Bayesian and frequentist algorithms accomplish in practice
and what can be proved about them.

Our goal remains partly achieved.
There are several places in generalized linear mixed
effects models where usual algorithms
impose a cost that is $\gg N^{3/2}$
per iteration making them completely infeasible.
We have removed all of those costs so that
iterations cost $O(N)$ each.
In our numerical examples we see a total cost
of $O(N)$ so that the number of iterations has
scaled as $O(1)$.  What remains is to find
more explanations of this phenomenon
and sufficient conditions to bound the iteration
count.

\section*{Acknowledgment}

We thank Stitch Fix for sharing some data with us
and are especially grateful to Bradley Klingenberg
and Sven Schmit.
This work was supported in part by
a National Science Foundation BIGDATA
grant IIS-1837931.

\bibliographystyle{chicago}
\bibliography{bigdata}

\appendix
\section{Some proofs}\label{sec:appendix}
\subsection{Proof of Lemma~\ref{lem:pert}}\label{sec:proof:lem:pert}
The first derivative of $T^{-1}$ with respect to $\eta$ is
\begin{align*}
\frac\mrd{\mrd\eta}T(\eta)^{-1}
&=-T(\eta)^{-1}T'(\eta)T(\eta)^{-1}
\quad\text{for}\quad T'(\eta) =\frac{\mrd}{\mrd\eta}T(\eta)
=\begin{pmatrix} 0 & B\\ B^\tran &0\end{pmatrix}.
\end{align*}
Higher order derivatives of $T(\eta)^{-1}$
simplify greatly because $T''(\eta)=0$.
For integers $k\ge1$ we find by induction that
\begin{align*}
\frac{\mrd^k}{\mrd\eta^k}T(\eta)^{-1}
&=
(-1)^{k}k!\bigl( T(\eta)^{-1}T'(\eta)\bigr)^kT(\eta)^{-1}\\
&=
(-1)^{k}k!
T(\eta)^{-1/2}
\Bigl(
T(\eta)^{-1/2}T'(\eta)T(\eta)^{-1/2}
\Bigr)^kT(\eta)^{-1/2}.
\end{align*}
Next, for
$B_* = A^{-1/2}BC^{-1/2}$,
we find that
$$T(0)^{-1/2}T'(0)T(0)^{-1/2}
=
\begin{pmatrix}
0 & -B_*\\
-B_*^\tran  & 0
\end{pmatrix}
$$
and then taking a Taylor series around $\eta=0$ we have
\begin{align}\label{eq:trapprox}
T(\eta)^{-1} =
T(0)^{-1}
+
T(0)^{-1/2}
\sum_{k\ge1}
\begin{pmatrix}
0 & -\eta B_*\\
-\eta B_*^\tran  & 0
\end{pmatrix}^k
T(0)^{-1/2}.
\end{align}
This Taylor series converges for $|\eta|<1/\rho$.

Therefore, for $0\le\eta<1/\rho$,
\begin{align*}
\tr(T(\eta)^{-1})
&= \tr(T(0)^{-1})+\tr\bigl( T(0)^{-1/2} E(\eta) T(0)^{-1/2}\bigr)\\
&= \tr\bigl(T(0)^{-1}(I+E(\eta)\bigr),\quad\text{for}\\
E(\eta) &=\sum_{k\ge1}
\begin{pmatrix}
0 & -\eta B_*\\
-\eta B_*^\tran  & 0
\end{pmatrix}^k.
\end{align*}
Recalling that $T(0)$ is diagonal,
the $k=1$ term does not contribute to the trace of the inverse.
That is also true for any every odd integer $k\ge1$.
It follows that
\begin{align*}
\tr(T(\eta)^{-1})&= \tr\bigl(T(0)^{-1}(I+E_2(\eta)\bigr),\quad\text{for}\\
E_2(\eta)
&= \sum_{k\ge1}
\begin{pmatrix}
0 & -\eta B_*\\
-\eta B_*^\tran  & 0
\end{pmatrix}^{2k}
= \sum_{k\ge1}
\begin{pmatrix}
\eta^2B_*B_*^\tran & 0\\
0 & \eta^2B_*^\tran B_*
\end{pmatrix}^{k}.
\end{align*}
Now we can write the difference $\tr(T(\eta)^{-1})-\tr(T(0)^{-1})$ as
$$
\sum_{k\ge 1}\eta^{2k}\Bigl(\tr\bigl(
A^{-1}(B_*B_*^\tran)^{k}\bigr)
+\tr\bigl(C^{-1}(B_*^\tran B_*)^{k}\bigr)\Bigr).
$$

\subsection{Proof of Lemma~\ref{lem:spectralbound}}
\label{sec:proof:lem:spectralbound}
First $\rho^2$ is no larger than the spectral radius of
$$
\begin{pmatrix}
0 & B_*\\
B_*&0
\end{pmatrix}^2
=
\begin{pmatrix}
B_*B_*^\tran & 0\\
0 & B_*^\tran B_*
\end{pmatrix}.
$$
The eigenvalues of this matrix are the same as those
of $B_*^\tran B_*$ and of
$$
B_*B_*^\tran = A^{-1/2}BC^{-1}B^\tran A^{-1/2}
$$
which has by similarity, the same eigenvalues as
$BC^{-1}B^\tran A^{-1}$.

Now $BC^{-1}$ is a nonnegative matrix and so
its largest eigenvalue is a real number no larger
than its largest row sum, which in this case is
$\max_i \sum_jB_{ij}/C_{jj}$.
Applying the same
to $B^\tran A^{-1}$ gives us a bound on $\rho^2$
that is the square of the claimed bound for $\rho$.

\subsection{Proof of Theorem~\ref{thm:eigvalso1}}
\label{sec:proof:thm:eigvalso1}
The $is$ element of $B_*B_*^\tran$ is
$$
(B_*B_*^\tran )_{is}=
\sum_j
\frac{\zij\wij}{\sqrt{\wid+\ssai}\sqrt{\wdj+\ssbi}}
\frac{\zsj\wsj}{\sqrt{\wsd+\ssai}\sqrt{\wdj+\ssbi}}.
$$
Therefore the
$i$'th diagonal element of $B_*B_*^\tran $ is
\begin{align*}
(B_*B_*^\tran )_{ii}=
\sum_j
\frac{\zij\wij^2}{{(\wid+\ssai)(\wdj+\ssbi)}}
\le\frac{1}{\wid}\sum_{j}\frac{\zij\wij^2}{\wdj}.
\end{align*}
With probability tending to $1$
$$
(B_*B_*^\tran )_{ii}\le
S^{2\psi}\bar\omega^2\frac1{\nid}\sum_j\frac{\zij}{\ndj}
\le
S^{2\psi}\bar\omega^2(1-\epsilon)^{-2}S^{\rho+\kappa-2}
\sum_j\zij
$$
by equations~\eqref{eq:gotnid} and~\eqref{eq:gotndj}.
Now $\sum_j\zij=\nid\le (\pup+\epsilon)S^{1-\rho}$
by equation~\eqref{eq:gotnid}.
Therefore with probability approaching one,
\begin{align*}
\tr( B_*B_*^\tran)
&
\le
\bar\omega^2(1-\epsilon)^{-2}(\pup+\epsilon)
S^{2\psi+\rho+\kappa-1}=o(R)
\end{align*}
because $R=S^\rho$ and $\kappa<1$
and we can choose any $\psi>0$.
Now none of the eigenvalues of $B_*B_*^\tran$
can be negative.
If there are $R^\alpha=S^{\alpha\rho}$
eigenvalues larger than $\delta$ then
$$
\tr( B_*B_*^\tran) > \delta
S^{\alpha\rho}.
$$
That cannot hold for large $S$ if
$$\alpha > \underline\alpha\equiv 1 +\frac{2\psi+\kappa-1}{\rho}<1$$
for small enough $\psi$.
Therefore any
$\alpha\in(\underline\alpha,1)$
satisfies the condition in~\eqref{eq:ratealpha}.

\subsection{Proof of Theorem~\ref{thm:trapproxtheorem}}
\label{sec:proof:thm:trapproxtheorem}
Let $B_*B_*^\tran$ have eigenvalues
$\lambda_1\ge\lambda_2\ge\cdots\ge\lambda_R\ge0$
with corresponding unit norm eigenvectors $v_i$.
Then
\begin{align*}
\err
&=
\tr\Bigl( A^{-1}\sum_{k\ge1}
\sum_{i=1}^R\lambda_i^kv_iv_i^\tran\Bigr)
=\sum_{k\ge1}\sum_{i=1}^R \lambda_i^k
\tr( v_i^\tran A^{-1}v_i).
\end{align*}
Now with probability tending to $1$ as $S \to \infty$,
$$
\tr( v_i^\tran A^{-1}v_i)\le
\lambda_{\max}(A^{-1}) = \max_s(\wsd+\ssai)^{-1}
\le S^\psi(1-\epsilon)^{-1}S^{\rho-1}
$$
because $\wsd\ge S^{-\psi}\min_i\nid\ge S^{-\psi}(1-\epsilon)S^{1-\rho}$ with overwhelming probability.
Therefore with probability tending to $1$ as $S \to \infty$
\begin{align*}
\err
&\le
S^{\psi+\rho-1}(1-\epsilon)^{-1}\sum_{i=1}^R\frac{\lambda_i}{1-\lambda_i}.
\end{align*}
Now choose the $\alpha<1$
provided by Theorem~\ref{thm:eigvalso1}
and suppose that for $\delta>0$
that fewer than $R^\alpha$ of the $\lambda_i$
are larger than $\delta$.
Under this event which has probability
tending to one,
\begin{align*}
\err & \le
S^{\psi+\rho-1}(1-\epsilon)^{-1}
\biggl(
\frac{\delta \lambda_1 R^\alpha}{1-\lambda_1}
+\frac{R}{1-\delta}\biggr).
\end{align*}
The second term above is
$O(S^{\psi+\rho-1}R)=o(R)$
because $\rho<1$ and we can choose any $\psi>0$.

To control the first term, we use Proposition~\ref{prop:lambda1}
to bound $1/(1-\lambda_1)$.
The bound there takes the form $\lambda_1\le (1+a)^{-1}(1+b)^{-1}$
for $a=S^{\rho-1}/[(\pup+\epsilon)\bar\omega \ssa]$
and $b=S^{\kappa-1}/[(\pup+\epsilon)\bar\omega \ssb]$.
Therefore
\begin{align*}
\frac{1}{(1/\lambda_1 - 1)}& \le \frac1{a+b+ab}
\le \min\Bigl(\frac1a,\frac1b\Bigr)\\
&=\min( S^{1-\rho}\ssa,S^{1-\kappa}\ssb)\bar\omega(\pup+\epsilon).
\end{align*}
It now follows that the first term in $\err$ is
$O(S^\psi R^\alpha)=o(R)$ because $\alpha<1$
and we can choose any $\psi>0$.

\subsection{Proof of Lemma~\ref{lemma:clubbingconvergence}}\label{sec:proof:lem:clubbingconvergence}

Our iterative algorithm is designed to minimize
$$
\pl(\beta,\bsa,\bsb)
= \sum_{ij}\zij{\wh W}_{ij} (z_{ij}-\xij^\tran\beta-\ai-\bj)^2
+\ssiha\Vert\bsa\Vert^2+\ssihb\Vert\bsb\Vert^2,
$$
where all $\wh W_{ij}>0$,  $\min(\ssiha,\ssihb)>0$,
the matrix $\cx$ with rows $\xij^\tran$
has full rank, and $z_{ij}$ are fixed numbers.
This quadratic function has a positive definite Hessian
and a unique global minimum
$(\beta_*,\bsa_*,\bsb_*)$.

We introduce $\pl(\beta,\bsa;\bsb)$ which for any $\bsb\in\real^C$
gives us a function of $\beta\in\real^p$ and $\bsa\in\real^C$.
Similarly, we introduce $\pl(\beta,\bsb;\bsa)$ where this time
$\bsa$ is the parameter while $\beta$ and $\bsb$ are
the function's arguments.
Given $\beta^{(k)}$, $\bsa^{(k)}$ and $\bsb^{(k)}$
we minimize $\pl(\beta,\bsa;\bsb^{(k)})$ over
$\beta$ and $\bsa$ to get
$\beta^{(k+\frac12)}$ and $\bsa^{(k+1)}$.
Then we minimize $\pl(\beta,\bsb;\bsa^{(k+1)})$ over
$\beta$ and $\bsb$ to get
$\beta^{(k+1)}$ and $\bsb^{(k+1)}$.
Both minimizations choose arguments that make their
respective gradients equal to zero.
Therefore
\begin{align*}
&\pl(\beta^{(k)},\bsa^{(k)};\bsb^{(k)})
-\pl(\beta^{(k+\frac12)},\bsa^{(k+1)};\bsb^{(k)})\\
&=
\frac12
\begin{pmatrix}
\beta^{(k)}-\beta^{(k+\frac12)}\\
\bsa^{(k)}-\bsa^{(k+1)}
\end{pmatrix}^\tran
H(\beta^{(k+\frac12)},\bsa^{(k+1)};\bsb^{(k)})
\begin{pmatrix}
\beta^{(k)}-\beta^{(k+\frac12)}\\
\bsa^{(k)}-\bsa^{(k+1)}
\end{pmatrix}
\end{align*}
using $H$ to denote the Hessian of the parameterized
function.
Considering both steps at once we get
\begin{align*}
\Delta^{(k+1)}&\equiv\pl(\beta^{(k)},\bsa^{(k)},\bsb^{(k)})
-\pl(\beta^{(k+1)},\bsa^{(k+1)},\bsb^{(k+1)})\\
&=\frac12
\begin{pmatrix}
\beta^{(k)}-\beta^{(k+\frac12)}\\
\bsa^{(k)}-\bsa^{(k+1)}
\end{pmatrix}^\tran
H(\beta^{(k+\frac12)},\bsa^{(k+1)};\bsb^{(k)})
\begin{pmatrix}
\beta^{(k)}-\beta^{(k+\frac12)}\\
\bsa^{(k)}-\bsa^{(k+1)}
\end{pmatrix}\\
&+\frac12
\begin{pmatrix}
\beta^{(k+\frac12)}-\beta^{(k+1)}\\
\bsb^{(k)}-\bsb^{(k+1)}
\end{pmatrix}^\tran
 H(\beta^{(k+1)},\bsb^{(k+1)};\bsa^{(k+1)})
\begin{pmatrix}
\beta^{(k+\frac12)}-\beta^{(k+1)}\\
\bsb^{(k)}-\bsb^{(k+1)}
\end{pmatrix}.
\end{align*}

Now $\Delta^{(k+1)}\ge0$ and
$$
\sum_{k=0}^\infty\Delta^{(k+1)}\le
\pl(\beta^{(0)},\bsa^{(0)},\bsb^{(0)})
-\pl(\beta_*,\bsa_*,\bsb_*)<\infty
$$
and so this sum converges.
Noting that $H(\beta,\bsa;\bsb)$ and $H(\beta,\bsb;\bsa)$ are independent of $(\beta,\bsa,\bsb)$ and strictly positive definite, we conclude that $(\beta^{(k)},\bsa^{(k)},\bsb^{(k)})$ converges.
At the limit point, the
gradients of both $\pl(\beta,\bsa;\bsb)$
and $\pl(\beta,\bsb;\bsa)$ must vanish
and so therefore the gradient of $\pl(\beta,\bsa,\bsb)$
also vanishes there.

\subsection{Results of the binary regression in Section~\ref{sec:stitchfix} }
\label{sec:stitchfixresults}

Table~\ref{tab:Stitch_fix_binary_regression} shows
coefficient estimates and standard errors
for  plain logistic regression and
a generalized linear mixed model
logistic regression for the Stitch Fix data in Section~\ref{sec:stitchfix}.
Logistic is estimated to be naive when
$\wh{\mathrm{SE}}_{\LR}(\hat\beta_\LR)<\wh{\mathrm{SE}}_{\glmm}(\hat\beta_\LR)$
and inefficient when
$\wh{\mathrm{SE}}_{\glmm}(\hat\beta_\LR)>\wh{\mathrm{SE}}_{\glmm}(\hat\beta_\glmm)$.
Estimates that are more than double their corresponding standard error
get an asterisk.
\newcolumntype{d}[1]{D{.}{.}{#1}}
\begin{table}[h!]
\caption{\label{tab:Stitch_fix_binary_regression}
Stitch Fix Binary
Regression Results}
\setlength{\tabcolsep}{4pt}

\begin{tabular}{l d{4} d{3.8} d{4.8} d{4} d{2.8}}
\toprule
& \multicolumn{1}{c}{$\hat{\beta}_{\logistic}$} & \multicolumn{1}{c}{$\widehat{\mathrm{SE}}_{\logistic}(\hat{\beta}_{\logistic})$} & \multicolumn{1}{c}{$\widehat{\mathrm{SE}}_{\glmm}(\hat{\beta}_{\logistic})$} & \multicolumn{1}{c}{$\hat{\beta}_{\glmm}$} & \multicolumn{1}{c}{$\widehat{\mathrm{SE}}_{\glmm}(\hat{\beta}_{\glmm})$} \\

\midrule
Intercept& 0.677^{*} & 0.022 & 0.048 & 0.584^{*} & 0.036\\
$\mathbb{I}\{\text {client fit profile missing}\}$ & 0.099^{*} & 0.015 & 0.025 & 0.127^{*} & 0.023 \\
 $\mathbb{I}\{\text {client fit profile loose}\}$ & 0.0155^{*} & 0.002 & 0.006 &0.208^{*} & 0.004 \\
 $\mathbb{I}\{\text {client fit profile oversize}\}$ & -0.0940^{*} & 0.010 & 0.019 & -0.0477^{*} & 0.015 \\
 $\mathbb{I}\{\text {client fit profile straight}\}$ & 0.125^{*} & 0.003 & 0.005 & 0.154^{*} & 0.004 \\
 $\mathbb{I}\{\text {client fit profile tight}\}$ & -0.229^{*} & 0.011 & 0.1297 & -0.250^{*} & 0.009 \\
 $\mathbb{I}\{\text {client edgy}\}$ & -0.042^{*} & 0.002 & 0.005 & -0.044^{*} & 0.003 \\
$\mathbb{I}\{\text {client boho}\}$ & 0.147^{*} & 0.002 & 0.006 & 0.194^{*}& 0.003\\
 $\mathbb{I}\{\text {client chest size}\}$ & -0.008^{*} & 0.0006 & 0.001 &-0.011^{*}& 0.001  \\
 $\mathbb{I}\{\text {client dress size}\}$ & -0.002^{*} & 0.0003 & 0.0007 &-0.0002 & 0.0005  \\
$\mathbb{I}\{\text {item fit profile missing}\}$ & -0.281^{*} & 0.024 & 0.188 & -0.079 & 0.151 \\
 $\mathbb{I}\{\text {item fit profile fitted}\}$ & 0.061^{*} & 0.004 & 0.089 & 0.277^{*} & 0.029 \\
 $\mathbb{I}\{\text {item fit profile loose}\}$ & 0.209^{*} & 0.003 & 0.082 & 0.038 & 0.029 \\
 $\mathbb{I}\{\text {item fit profile oversize}\}$ & -0.066^{*} & 0.016 & 0.135 & -0.020 & 0.111 \\
 $\mathbb{I}\{\text {item fit profile tight}\}$ & 0.186 & 0.0098 & 0.431 & 0.251 & 0.359 \\

 $\mathbb{I}\{\text {material missing}\}$ & -0.203^{*} & 0.022 & 0.190 & -0.227 & 0.135 \\
 Acrylic & -0.187^{*} & 0.005 & 0.060 &-0.206^{*} & 0.031                     \\
 Angora & -0.305^{*} & 0.024 &  0.222 &  -0.279 & 0.151\\
Cashmere & 0.367^{*} & 0.057 & 0.462 & -0.744& 0.399     \\
 Cotton & -0.246^{*} & 0.004 & 0.060& -0.245^{*}  & 0.028              \\
 Faux Fur & 0.321 & 0.250 & 0.516 & 0.547 & 0.479                 \\
 Fur & -0.459^{*} & 0.128 & 0.489 & -0.619 & 0.480    \\
 Linen & -0.494^{*} & 0.025 & 0.182 & -0.374^{*}&0.163\\
 Modal & 0.035^{*} & 0.007 & 0.262&-0.057 &0.080                         \\
 Nylon & 0.096^{*} & 0.013 & 0.188 &0.070&0.108\\
 Patent Leather & -0.789^{*} & 0.110&0.454 & -0.531&0.379   \\
 Pleather & -0.215^{*} & 0.020 & 0.194 & -0.121&0.130 \\
 PU & 0.390^{*} & 0.042 & 0.0462 & 0.509&0.0351\\
 Rayon & -0.033^{*} & 0.002 & 0.068 & -0.013&0.020 \\
 Silk & -0.041^{*} & 0.009 & 0.133 & 0.053&0.068 \\
 Spandex & 0.025 & 0.050 &0.355  & 0.210 &0.335  \\
  Tencel & -0.107^{*} & 0.025 & 0.181 & -0.041&0.158   \\
 Viscose & -0.081^{*} & 0.008 & 0.086 & -0.084&0.054   \\
Wool & -0.217^{*} & 0.024 & 0.235&-0.194&0.150\\
\bottomrule
\end{tabular}

\end{table}
For the Stitch Fix data we obtained
$\hat{\sigma}_{A}^{2} = 0.68$ (customers),
$\hat{\sigma}^{2}_{B} = 0.21$ (items).

\end{document}